\documentclass[12pt,a4paper]{article}
\usepackage{amsmath}
\usepackage{amssymb}
\usepackage{amsthm}
\usepackage{here}
\usepackage{graphicx}
\usepackage{verbatim}

\newtheorem{proposition}{\bf Hypothesis}

\theoremstyle{definition}

\input{tcilatex}

\begin{document}

\title{The economics of minority language use: theory and empirical evidence
for a language game model}
\author{Stefan Sperlich\thanks{
Geneva School of Economics and Management, Universit\'{e} de Gen\`{e}ve, Bd
du Pont d'Arve 40, CH-1211 Gen\`{e}ve, Suisse. E-mail address: \texttt{\
stefan.sperlich@unige.de}.} \and Jos\'{e}-Ram\'{o}n Uriarte\thanks{
{Corresponding author:} University of the Basque Country, Departamento de
Fundamentos del An\'{a}lisis Econ\'{o}mico I, Avenida Lehendakari Aguirre
83, E-48015 Bilbao, Basque Country-Spain. E-mail address: \texttt{\
jr.uriarte@ehu.es} } }
\maketitle

\begin{abstract}
\noindent Language and cultural diversity is a fundamental aspect of the
present world. We study three modern multilingual societies -the Basque
Country, Ireland and Wales- which are endowed with two, linguistically
distant, official languages: $A$, spoken by all individuals, and $B$, spoken
by a bilingual minority. In the three cases it is observed a decay in the
use of minoritarian $B$, a sign of diversity loss. However, for the \textit{%
Council of Europe} the key factor to avoid the shift of $B$ is its use in
all domains. Thus, we investigate the language choices of the bilinguals by
means of an evolutionary game theoretic model. We show that the language
population dynamics has reached an evolutionary stable equilibrium where a
fraction of bilinguals have shifted to speak $A$. Thus, this equilibrium
captures the decline in the use of $B$. To test the theory we build
empirical models that predict the use of $B$ for each proportion of
bilinguals. We show that model-based predictions fit very well the observed
use of Basque, Irish, and Welsh.\footnote{%
This joint work started while Jos\'{e}-Ram\'{o}n Uriarte was visiting the
Humbolt-Universit\"{a}t zu Berlin and the University of Geneva. He would
like to thank the generous hospitality of both universities. Financial
support provided by the Basque Government, by the Ministerio de Econom\'{\i}%
a y Competitividad (ECO2012-31626, ECO2016-76203-C2-1-P), and by the
Santander Financial Institute are gratefully acknowledged.}


\end{abstract}

\section{Introduction}

\label{sec-Intro}

\begin{quote}
\textit{As a source of exchange, innovation and creativity, cultural
diversity is as necessary for humankind as biodiversity is for nature}.
UNESCO (2002), Article 1.
\end{quote}

The present work deals with the issue of preserving language and cultural
diversity in highly developed multilingual societies having two \textit{%
linguistically distant}\footnote{%
See Ginsburgh and Weber (2011) for a survey on how to compute the distance
between two languages; see also Desmet et al.\ (2009).} official languages: $%
A$, denoting the language spoken by all individuals, and $B$, denoting the
one spoken by a \textit{bilingual minority}. Some examples of advanced
multilingual societies with distant official languages and bilingual
minorities are the Basque Country, Ireland and Wales. They are endowed with
enough resources to design and implement well-articulated language policies
in favour of $B$, which means, among other things, that they have a public
bilingual educational system (see Mercator, 2008, 2014, 2016), media, and
markets for cultural goods related with language $B$. The idea is that these
multilingual societies will set a kind of benchmark into the set of
societies with threatened languages contemplated in Fishman (2001). The
transmission of bilingual cultural traits in these multilingual societies
works, fundamentally, through the public educational system (parents send
their children to the public system to receive a bilingual education, so
that they learn the two official languages $A$ and $B$). Thus, the
institutions in charge of the cultural transmission generate a continuous
flow of potentially bilingual individuals.

Cultural transmission was formally studied by Cavalli-Sforzza and Feldman
(1981) by means of an evolutionary biology-based model, in a society with
two cultural traits distributed in the population. Their model produces two
stable equilibria, each allowing the survival of just one trait, and one
unstable equilibrium where the two traits are present. Thus, the model did
not capture well the diversity observed in certain societies, particularly
in those with minoritarian cultural groups. Bisin and Verdier (2001, 2010),
extended the Cavalli-Sforzza and Feldman model avoiding the boundary
equilibria, and allowing for a stable interior equilibrium where the two
traits can coexist in the population state. The present work adopts an
evolutionary game-theoretic approach, free of biological constraints, to
study the population dynamics of cultural traits represented by the two
official languages, $A$ and $B$\footnote{%
We will model the population dynamics by means of the replicator dynamics.
Even though the origin of the replicator equation is in biology (Maynard
Smith, 1982), it has been shown that, in socio-economic contexts, it can be
obtained from diverse trial-and-error learning procedures (see, among
others, Binmore et al.,1995, Cabrales, 2000, Cabrales and Uriarte, 2013, and
Schlag, 1998).}. However, we do not study language and culture transmission.
Instead, we concentrate our analysis on the weakening of bilingual cultural
traits that are observed in the data about the use of Basque, Irish and
Welsh.

The data show that despite the existing proportions of speakers in those
languages, the use of $B$ is lagging far behind. This seems to be a paradox
in societies having a strong policy in favour of $B$. Why is it that having
the political system -- that is, elected governments promoting the use of $B$
-- the legal instruments to facilitate its use\footnote{%
The \textit{Law of Normalisation of Basque's Use} of 1982; the \textit{Welsh
Language Measure} of 2011; and the recognition by \textit{Constitution of
the Republic of Ireland} that Irish as the first official language.},
resources, bilingual educational systems, and the bilinguals' preference for 
$B$, that the same people make a low use of $B$? Why is it that, knowing the 
\textit{Council of Europe's }\footnote{\textit{European Charter for Regional
or Minority Languages} www.coe.int/t/dg4/education/minlang} advice that 
\textit{the key factor to avoid a minority language shift is that }$B$%
\textit{\ should be used in all social domains}, speakers of $B$ often use $A
$ when they interact between them, putting at risk the existence of $B$ and
its related culture?

The purpose of this work is to analyse the source of this paradox which,
adopting the \textit{Council of Europe}'s view, must be the process by which
bilinguals decide the language to use in each interaction. Our view is that
the survival of language $B$ depends, essentially, on the bilingual
individuals, i.e.\ on how intense is their preference for using $B$ in their
interactions.

Without denying the links with the cultural transmission literature, our
work is closer to the literature about endangered minority languages and
cultural diversity. This is a topic widely studied by sociolinguists (e.g.
Fishman 1991, 2001; see a description of language death process in Crystal,
2000 ), and by physicists and mathematicians working in complex systems
(Abrams and Strogatz, 2003 is the seminal work; see, among others, Stauffer
and Schulze, 2005, Minett and Wang, 2008, Sol\'{e} et al., 2010; this
approach is surveyed in Patriarca et al., 2012). These two schools frame the
process of threatening and eroding cultural diversity in the context of
language competition dynamics. That is, natural languages, with their speech
communities, are viewed as competing for speakers, very much like firms
compete for market shares\footnote{%
As remarked by Nelde (1995) "neither contact nor conflict can occur between
languages; they are conceivable only between speakers of languages and
between the language communities".}. Surprisingly, this competitive
situation has received little attention from economic theorists and from
game theory practitioners\footnote{%
\textquotedblleft The lack of economic analysis of the natural language that
characterizes human economic behaviour is certainly a large and visible
hole\textquotedblright . Anderlini and Felli (2004). Notable exceptions are
Rubinstein (2000) and Selten and Pool (1991). Grin et al.\ (2010) and
Ginsburgh and Weber (2011) are recent surveys of the research developed in
the economics of language.}. Our framework uses notions and analytical tools
of behavioural economics, to study the language competition between $A$ and $%
B$, avoiding some undesirable features of both sociolinguistics and the
formal and complex systems approach. In sociolinguistics it is the lack of
formalization that may help to throw light into the complexities of language
competition. Physicists and mathematicians have yet to extend their models
to allow for adaptive decision makers giving rise to more realistic
population dynamics and behavioural results. These drawbacks are eluded by
our model.

We highlight two factors that condition the bilinguals' language choices.
One is the \textit{information} that bilingual individuals have about the
linguistic types (bilingual or monolingual) of the interactive partners.
Very likely, that information was perfect in the past; in small, traditional
communities with low immigration and mobility. However, in modern societies
bilinguals are spread all over the country, particularly in urban areas%
\footnote{%
More than half of the Basque speakers live in the six most populated towns
of the Country. Similarly, 75.7 \% of Welsh speakers live in towns of more
than 100,000 inhabitants. Irish speakers outside the Gaeltacht are a
minority, raised and educated through Irish, living in urban areas.}, where
interactions are frequently anonymous, and the linguistic type tends to be
private information. Thus, bilinguals must often choose the language of
interaction without (perfect) information on the interlocutor's linguistic
type. The other factor is that in these societies there is what is called a 
\textit{language contact situation }(Nelde, 1995 and Winford, 2003) which
means that the languages spoken by bilinguals are being altered,
particularly $B$, due to the constant interaction between bilinguals and
monolinguals using the majority language $A$.

We assume that bilinguals choose the language of interaction through an
extended version of the \textit{Language Use Game} (LUG) which is based on a
game introduced by Iriberri and Uriarte (2012) (IU henceforth). The LUG is a
game with imperfect information of linguistic types, in which bilinguals may
choose between two pure strategies, either $Reveal(R)$ the bilingual type or 
$Hide(H)$ it. Imperfect information and language contact leave a tight room
to satisfy the preference for the use of $B$. Under these circumstances,
bilinguals are bound to make frequent language choices, giving rise to
unconscious patterns of choice behaviour. Thus, we assume that the
bilinguals play continuously the LUG, and learn by a trial-and-error
adjustment process, modelled by the one-player replicator dynamics attached
to the LUG. We show that this dynamic process converges into an interior 
\textit{evolutionary stable} \textit{Nash equilibrium}. This kind of
equilibria are considered as conventions (Weibull, 1995). Therefore, in the
context of the LUG, this equilibrium would be a \textit{linguistic convention%
} in which the bilingual population is partitioned in two groups. One group
is composed of those who choose $R$; they will always speak $B$ with any
bilingual. The other group is composed by by those who choose $H$: they
speak $A$ between them. Thus, this equilibrium provides a theoretical
explanation of the observed decline in the use of $B$. To test empirically
this theory, we first make the Nash equilibrium to depend on the proportion
of bilinguals, $\alpha \in (0,1)$, and then we build econometric models that
predict, for each $\alpha $, the rate of use of $B$ in equilibrium. Finally,
the predictions are compared with the observed use of Basque, Irish, and
Welsh. The main finding is that the predictions fit well the observed data.
This is confirmed when testing them with nonparametric (i.e.\ model-free)
data fits. Moreover, where data are available over time we see that the
functions change quite smoothly.

The rest of the paper is structured as follows. In Section \ref{sec-KEdef}
we introduce the general framework of the LUG. Section \ref{sec-Nashequ} we
extend the LUG introducing  behavioural assumptions to build a Nash
equilibrium function that depends on the proportion of bilingual speakers.
In Section \ref{sec-empiric} we test the model for the Basque Country,
Ireland and Wales. Section \ref{sec-concl} concludes, summarizing results
and pointing out policy measures. Section 6 contains the Appendix.

\section{The Theoretical Framework}

\label{sec-KEdef}

We will be dealing with economically advanced multilingual societies having
two official languages, denoted $A$ and $B$. We assume that in those
societies there are essentially two linguistic groups: the monolingual
speakers, who speak only $A$, and a minority of bilingual speakers who
speak, with similar skills, $A$ and $B$\footnote{%
It should be clear from the outset that we are referring only to bilingual
speakers in the two `internal' official languages $A$ and $B$.}. Let $\alpha
>0$ denote the proportion of bilingual speakers in the society, and $%
1-\alpha $ the proportion of monolinguals. We assume $\alpha <1-\alpha $,
and $\alpha \in (0,1)$.\footnote{%
The boundary numbers 0 and 1 are hardly representatives.} The first
assumption we make says that communication takes place in a single language,
either $A$ or $B$.

\textbf{A.1. The Languages} \textit{with official status, }$A$\textit{\ and }%
$B$\textit{, are linguistically distant so that successful communication is
only possible when the interaction takes place in one language.}

Individuals are therefore assumed to view $A$ and $B$ having the same
status. We show below that \textbf{A.1} is consistent with the data. Passive
bilinguals, i.e.\ those who understand $B$ but cannot speak it, are,
therefore, not considered (or can be treated as monolinguals; see empirical
part). When a monolingual interacts with a bilingual, they necessarily use $%
A $. Thus, $B$ is spoken only if two bilinguals meet, and at least one
signals the desire to speak in $B$. Hence, language choice is not trivial.

Some examples of societies with official languages satisfying \textbf{A.1}
are the Basque Country, Ireland and Wales. Basque is a pre-indoeuropean
language in contact with the two Romance languages, Spanish (or Castellano)
and French. Irish and Welsh, two Celtic languages, are in contact with the
Germanic language English. These are competitive multilingual societies,
both in the economic domain, and in the linguistic domain, with sufficient
resources to implement well articulated language policies to protect and
transmit $B$ through the educational system. Thus, we will analyse
theoretically and empirically the factors that might affect {the use} of $B$
in multilingual societies which are highly developed economically. Our view
is that these societies form a kind of {benchmark multilingual societies} in
the set of all societies with threatened languages satisfying \textbf{A.1}.
If minority languages will be shifted there, this will even be more so in
the less developed multilingual societies.

We may measure the use of $B$ in different ways. We can measure the
frequency of use of $B$ (say, daily, weekly, less often, never) as it is
done in the censi and surveys about the Irish and Welsh languages. Or we may
measure the use of $B$ in the streets, as it is done for Basque. For Irish
and Welsh we use observations about the \textit{Daily Use} ($DU$), for
Basque we have the observations of \textit{Street Use} ($KE$); see Appendix
6.1 for details, and how $\alpha $, $KE$ and $DU$ are defined.

\begin{table}[h]
\caption{Evolution of percentage of bilinguals $\protect\alpha $ and Street
Use or Daily Use for Basque, Irish and Welsh. For Basque, $\protect\alpha $
and $KE$ was not always measured in the same year; e.g.\ 1991/3 indicates
that $\protect\alpha $ is from 91, $KE$ from 93. For Welsh, the surveys were
executed over longer periods.}
\label{tab1Basque}
\begin{center}
\begin{tabular}{lll|lll|lll}
\multicolumn{3}{c|}{Basque} & \multicolumn{3}{c|}{Irish} & 
\multicolumn{3}{c}{Welsh} \\[2mm] \hline
Year & $100\alpha$ & $100KE$ & Year & $100\alpha$ & $100DU$ & Years & $%
100\alpha$ & $100DU$ \\ \hline
1991/3 & 22.3 & 11.8 &  &  &  &  &  &  \\ 
1996/7 & 24.4 & 13.0 & 1996 & 41.10 & 10.16 &  &  &  \\ 
2001 & 25.4 & 13.3 & 2002 & 41.88 & 09.05 &  &  &  \\ 
2006 & 25.7 & 13.7 & 2006 & 40.83 & 02.10 & 2004-06 & 20.5 & 13.0 \\ 
2011 & 27.0 & 13.3 & 2011 & 40.60 & 02.15 &  &  &  \\ 
2016 & 28.4 & 12.6 &  &  &  & 2013-14 & 19.0 & 10.1%
\end{tabular}%
\end{center}
\end{table}

Clearly, $KE$ and $DU$ will be modelled differently, but we suppose that
both result from the same language game, and are functions of the same
factors. Table \ref{tab1Basque} shows $\alpha $, $KE$ or $DU$, for the case
of Basque, Irish, and Welsh for different years. Some features of the
numbers are worth mentioning. While $\alpha $ is either steadily increasing
(for Basque and Welsh) or on a very high level (Irish), the use of $B$ is
lagging behind, stagnating or even going down. Note that we observe a
structural break of $DU$ for Irish between 2002 and 2006. This may be due to
the fact that only since 2006 the daily use of Irish is clearly defined as
\textquotedblleft used outside the education system\textquotedblright . In
Table 1 only the aggregates are given; for the empirical study we will use
the data taken on small area level exhibiting the joint variation of $\alpha 
$ and language use.

The bilingual population in each country displays an empirical regularity
that is commonly observed in social groups: that the bilingual individuals
tend to behave similarly (not identically, as we will see below in the
equilibrium analysis of section 2.1.4) with respect to the use of language $%
B $. This universal regularity is, for Manski (1997), the result of a type
of social process called \textit{anonymous endogenous interactions}. It is a
social process in which the behaviour of an individual varies with the
distribution of behaviour in the reference group. The endogeneity of
interactions will be formalized below by means of the \textit{replicator
dynamics equation } associated to the game played by the bilingual
population.

\subsection{A Game- Theoretic Model of Language Choice}

\label{sec-model}

We seek to understand the data of $B$ use. To this end, we model the
strategic manner by which the members of bilingual population choose, at the
beginning of an interaction, the language to use with the interactive
partners. Bilinguals are assumed to choose \textit{only} pure strategies,
and adjust their choices endogenously by trial-and-error learning
procedures. We will show that this adaptive process converges into a Nash
equilibrium with the properties of a linguistic convention.

\subsubsection{Anonymous Interactions in Modern Multilingual Societies}

A feature of modern societies is the mobility, both social and geographical,
of the work force. As a consequence, interactions become increasingly
anonymous. In the benchmark multilingual societies, the spread of bilinguals
across the layers of society would mean that frequently bilinguals
participate in interactions without recognizing each other, because the
linguistic types (bilingual or monolingual) have become private information%
\footnote{%
In a very different context, but related with natural language issues, see
how language competence is made private information in Blume and Board
(2013).}.

A key source of that imperfect information, that enhances the one derived
from modernization, is the behaviour of the bilingual individuals
themselves. For Nelde (1995) "contact between languages always involves an
element of conflict". More specifically, if a monolingual is addressed in $B$%
, or observes an interactive partner who is signalling the desire to speak
in $B$, then the monolingual is forced to reveal his type, and must confess
his lack of knowledge of $B$. However, the need of communication will
frustrate the bilingual's desire, who, by \textbf{A.1}, will also be forced
to switch, sometimes reluctantly, to $A$. This might create a tension
between the interactants (not free of political undertones, in certain
cases) that could interfere the final goal of the interaction. Thus, sending
signals conveying the intention to speak in $B$ might be harmful for both
sides\footnote{%
We could observe the subtleties of this potential conflict by means of 
\textit{linguistic politeness theory} (see Brown and Levinson, 1987). For a
detailed discussion on this, see Uriarte (2017).}. Bilinguals experience
frequently this kind of interaction. However, given their linguistic
flexibility, a growing proportion of them tend to avoid the potential
tensions with unknown interlocutors, and choose to speak $A$, mistakenly
arguing that it is not \textit{polite} to start with $B$ (see Amorrortu et
al., 2009 for the case of Basque). Therefore, the politeness induced choice
behaviour preserves the imperfect information about linguistic types.
Further, this behaviour is reinforced if the probability of meeting a
monolingual is much higher than meeting a bilingual; that is, because $%
1-\alpha >\alpha $.

We should also mention a new breed of bilinguals, called \textquotedblleft
new speakers\textquotedblright , who have learnt the minority language by
means other than the family transmission, essentially through immersion
learning models in the educational system. These newcomers are inclined to
deny themselves as \textquotedblleft authentic\textquotedblright\ and
legitimate speakers of $B$ as compared to the native speakers, even though
they are fully competent in language $B$. They feel insecure with $B$, and
tend to not use it outside the school, thus, hiding their bilingual type to
interlocutors. See Ortega and al.\ (2015) for Basque, O'Rourke and Ramallo
(2011) for Irish, and Selleck (2018), for Welsh.

Additional sources of anonymity, among others, are: (a) the continuous
contact and interaction with speakers of language $A$ erase some revealing
signals of $B$ speakers, such as the accents. The contact situation makes
both bilinguals and monolinguals to have a similar accent, which is shaped
by the dominant language\footnote{%
For example, in the Spanish side of the Basque Country, bilingual and
monolingual people have similar Spanish accent, while in the French side
they have a French accent.}; (b) Second and third generations of immigrants
learn language $B$ in the public educational system. Thus, differences in
ethnic features, if any, hardly reveal the linguistic type. Thus the
anonymity of interactions is the result of a combination of several sources.
We are then led to conclude that it is realistic to assume that in the
benchmark multilingual societies linguistic types are, often, \textit{%
private information}\footnote{%
To avoid this, Welsh speakers may wear the Cymraeg badges of the Welsh
Language Comission. There are also badges for Welsh learners, as well as
lanyards and email footer, etc. This does not exist in the Basque Country,
probably because it would be politically controversial. Signs announcing
that Basque is spoken in official buildings, banks and shops are fairly
common though.}.

\textbf{A.2. Imperfect information:} \textit{The participants of an
interaction do not have, ex-ante, any information about the linguistic type
(bilingual or monolingual) of each other. They only know the average
proportion of bilingual and monolingual speakers, }$\alpha $ \textit{and} $%
(1-\alpha )$ \textit{respectively, of the society and that }$0<\alpha
<(1-\alpha )$.\footnote{%
We assume that bilinguals know the average proportion $\alpha $ of
bilinguals in the society because a summary of sociolinguistic statistics
are periodically published, and talked about in the media. Note that Catalan
is a minority language inside the Spanish State, but inside Catalunya, out
of a population of 7,496,27, roughly $\alpha =76\%$ speak Catalan and 86\%
understand it. Further, since \textbf{A.1} is not satisfied between
Castellano and Catalan, and there is common knowledge that `almost' all
individuals speak or understand Catalan, there is no hindering of the flow
of a conversation nor any tension caused by starting in Catalan with unknown
interlocutors. Hence, neither \textbf{A.1} nor \textbf{A.2 } apply.
Similarly, the theories developed for Catalan would not apply here, see
IDESCAT.}

\textbf{Remark:} One may argue that it is very likely that the street use of 
$B$ occurs also in assortative matching. But any econometric model (such as
the ones we estimate later) is a trade-off between a simplifying economic
model and reality. We may account for assortative matchings by introducing a
parameter $p$ denoting the proportion of random matches and $1-p$ that of
assortative matchings of bilinguals. We did this in an earlier version of
the work, but the additional parameter $p$ did neither lead to deeper
insights nor to better fits\footnote{%
Say, by introducing the parameter $p$, the empirical Model 2, that predicts
the street use of B, described below, would be $p\alpha ^{2}(x^{\ast
})^{2}+(1-p)\alpha ^{2}$.}.

\subsubsection{Linguistic Preferences and Payoffs}

The fact that language $B$ has become official reflects not only that
linguistic rights of the minority language speakers are respected, but also
that bilinguals manifest a preference for communication in language $B$. We
may assume that this linguistic preference could be weighted by a proper
choice of payoffs. Payoffs are assumed to represent net communication
benefits (see Selten and Pool, 1991, and Selten and Warglien, 2007). A
monolingual speaker cannot make language choices; thus, assuming that every
individual has the same language competence and communication skills in $A$,
a monolingual speaker will get a sure payoff $n$. Since language choices are
made under imperfect information, a bilingual may choose voluntarily
language $A$. In that case, we assume that he gets the payoff $n$. Bilingual
speakers will get the maximum payoff, $m$, when they coordinate in their
preferred language $B$. However, $(n-c)>0$ would be the payoff to a
bilingual speaker who, having chosen $B$, is matched to a monolingual, and
therefore is forced to switch to $A$. Then $c$ denotes the frustration cost.
In sum:

\textbf{A.3. } \textbf{Language preferences and Payoffs:} \textit{bilingual
speakers prefer to use language} $B$. \textit{Further,} \textit{it holds }$%
m>n>c>0$\textit{, and that the frustration cost is smaller than the weighted
benefits, i.e. } $c<(m-n)\frac{\alpha }{(1-\alpha )}=b(\alpha )$.

\subsubsection{The Language Use Game (LUG)}

Under the assumptions \textbf{A1-A.3}, bilinguals, at the beginning of an
interaction, may take a decision about the language to be used in the
interaction. An ordinary conversation has a sequential nature. Therefore,
the choice of language is better described by a game in extensive form, such
as the game of IU (Iriberri and Uriarte, 2012). For empirical purposes we
present now its strategic form (which will be extended in Section 3). Under
the assumptions \textbf{A.1-A.3}, the bilinguals' language strategic
behaviour is captured by the following pure strategies of bilinguals:

$\mathbf{R}$: \textit{Reveal always your type, so that you will speak B
whenever you meet a bilingual.}

$\mathbf{H}$: \textit{Hide your type, and reveal it only when matched with
an R-player. That is, speak A, and switch to B only when you are addressed
in B.}

IU show the sequential nature of these two strategies. As a speaker (i.e. as
the interactant who starts the conversation) or as listener (i.e. the one
who comes after the speaker) an $R$ chooser will try to use $B$ (sending
signals of wanting to use $B$) \textit{to discover} the interlocutor's
linguistic type, so that $B$ will be used when he meets another bilingual%
\footnote{%
When an interlocutor is addressed in $B$, he would feel forced, by \textbf{%
A.1 }and\textbf{\ A.3}, to confess his linguistic type (monolingual or
bilingual). A monolingual might feel unconfortable in this situation.}. In
this event he will get the maximum pay-off $m$; but in the event of meeting
a monolingual he gets $(n-c)$. An $H$ player will choose $A$ in the role of
speaker. Whereas, as a listener, he will answer in the language used by the
speaker, either $A$ or $B$. Therefore an $H$ chooser avoids the cost $c$,
and when he meets an $R$ chooser he will get $m$. The essence of $H$ is that
it reinforces the imperfect information. When two bilinguals who choose $H$
meet, they will speak in $A$. Language $B$ will be spoken whenever two
bilinguals meet, and, at least, one of them chooses $R$. An $H$ chooser
behaves like a monolingual in disguise unless he is discovered by an $R$
chooser. Clearly $H$ is less demanding than $R$ (recall the arguments for
the imperfect information assumptions).

There are two states of nature: \textit{Bilingual} and \textit{Monolingual}.
When the \textit{Bilingual} state is realized, two bilingual agents meet and
play the game described by the two-by-two matrix on the left side of Figure %
\ref{graphik1}. Each cell in this matrix lists the pay-off to the row and
column players in lower case letters ($m$ or $n$), and, for each strategy
combination, the resulting language of the interaction, shown in capital
letters ($A$ or $B$). When the \textit{Monolingual} state is realized the
bilingual meets a monolingual and, depending on the strategy chosen, gets
the pay-off shown by the column on the right side of Figure \ref{graphik1}.
In this state, the language of conversation is $A$, regardless of the
bilingual's strategy.\footnote{%
We skip the matching of two monolingual speakers since it is irrelevant, and
only show the bilingual's payoffs. When looking at the language use data,
the matches of monolinguals are taken into account, as they are a part of
our observations.} At the start of the game a state is realized. Bilinguals
do not know the state when they choose a pure strategy. They hold beliefs
assigning probability $\alpha $ to the state \textit{Bilingual} and
probability $1-\alpha $ to the state \textit{Monolingual}. Therefore, a
bilingual expects to meet another bilingual with probability $\alpha $, and
a monolingual speaker with probability $1-\alpha $. This comes from
assumption \textbf{A.2}.

\begin{figure}[tbh]
\begin{center}
\begin{tabular}{llllll}
Probability & $\alpha $ &  &  & Probability & $1-\alpha $ \\ 
& \textbf{R} & \textbf{H} &  &  &  \\ \cline{2-3}\cline{6-6}
\textbf{R} & \multicolumn{1}{|l}{$m,m;B$} & \multicolumn{1}{|l}{$m,m;B$} & 
\multicolumn{1}{|l}{} & \textbf{R} & \multicolumn{1}{|l|}{$n-c;A$} \\ 
\cline{2-3}\cline{6-6}
\textbf{H} & \multicolumn{1}{|l}{$m,m;B$} & \multicolumn{1}{|l}{$n,n;\ \ A$}
& \multicolumn{1}{|l}{} & \textbf{H} & \multicolumn{1}{|l|}{$n;\ \ \ \ \ A$}
\\ \cline{2-3}\cline{6-6}
&  &  &  &  &  \\ 
State & Bilingual &  &  & State & Monolingual%
\end{tabular}%
\end{center}
\caption{The Language Use Game.}
\label{graphik1}
\end{figure}

\subsubsection{The Language Use Game as a Population Game}

Assumptions \textbf{A.1} and \textbf{A.2 }joint with the close contact with
monolinguals capture the conditions under which bilinguals must make their
language choices. Then the bilinguals' linguistic preferences (\textbf{A.3})
lead them to make frequent language choices, aiming to get an efficient
communication and the coordination in their preferred language $B$.
Theoretically, we may view the LUG not as a one-shot game, but as a game
played continuously by bilinguals.\footnote{%
The game is played through pairwise random matches. That is, pairs of
individuals are repeatedly drawn at random to play the game through which
the language of interaction is determined.} In this manner, we have
converted the LUG into a game played by the bilingual population.

Bilinguals learn by a trial-and-error adjustment process, based on imitation
and interactive learning (i.e. endogenous interactions). At each moment of
time there is a population state $(N_{R}/N,N_{H}/N)$, where $N$ denotes the
total number of bilinguals (we suppress time in the notation), and $N_{i}/N$
represents the share of bilinguals playing pure strategy $i$, $(i=R,H)$. The
continuous play of the LUG produce changes in the population state and in
the payoffs to the pure strategies, resulting in a population dynamics in
continuous time, the one-player population \textit{replicator dynamics}
attached to the LUG (see Appendix 6.2). The replicators are the pure
strategies $R$ and $H$. This version of the replicator dynamics assumes that
bilinguals will choose \textit{only} pure strategies. Let $x$ denote the
proportion of bilingual speakers who choose to play $R$. Under \textbf{A.1}-%
\textbf{A.3}, and a given $\alpha \in (0,1)$, the bilingual player
population`s adjustment process converges into the unique evolutionary
stable equilibrium of the LUG, denoted $x^{\ast }$. This is the only
asymptotically stable equilibrium in the associated one-population
replicator dynamics: 
\begin{equation}
x^{\ast }=1-\frac{(1-\alpha )c}{\alpha (m-n)} \ , \ x^{\ast }\in (0,1).
\label{eqn(1)}
\end{equation}%
There are two additional Nash equilibria, $(R,H)$ and $(H,R)$, both
unstable, and the language of interaction in both equilibria is $B$, see
also Appendix 6.2.

In social or economic contexts, an evolutionary stable equilibrium, such as $%
x^{\ast }$, can be thought of as a {convention} (Weibull, 1995). Thus, the
population dynamics attached to the LUG converges into a {linguistic
convention}.\footnote{%
The one-shot play analysis of \textit{standard game theory} will not capture
the strong stability features of $x^{\ast }$ which are relevant in our
analysis.} This linguistic convention takes the form of a partition of the
bilingual population in two groups. The group $Nx^{\ast }=N_{R}^{\ast }$
consists of all the bilinguals who have chosen the pure strategy $R$ in
equilibrium. The other group, $N(1-x^{\ast })=N_{H}^{\ast }$, consists of
all the bilinguals who have chosen the pure strategy $H$ in equilibrium.%
\footnote{%
Hence, the equilibrium $(x^{\ast },1-x^{\ast })$, instead of being
interpreted as a randomization performed by each and every bilingual, it is
interpreted as a population state $(N_{R}^{\ast }/N,N_{H}^{\ast }/N)$. They
are both equivalent; see Weibull (1995).}

Any member of $N_{R}^{\ast }$ will speak in $B$ whenever he meets a
bilingual. However, when a bilingual of the group $N_{H}^{\ast }$ meets
another bilingual of the same group, they will speak in $A$. They will use $%
B $ only when they encounter a bilingual of the $N_{R}^{\ast }$ group. The
theoretical finding is described in the following proposition:

\textbf{Proposition}:\textit{\ Under the assumptions A1-A.3, and the
exogenously given parameters }$\alpha \in (0,1),m,n$\textit{\ and }$c$%
\textit{, the population dynamics (or equivalently, the language use
population dynamics) attached to the LUG converges into an evolutionary
stable mixed strategy Nash equilibrium, }$(x^{\ast },1-x^{\ast })$\textit{\ }%
$=(N_{R}^{\ast }/N,N_{H}^{\ast }/N)$\textit{. This equilibrium is a
linguistic convention which, since }$x^{\ast }>0$,\textit{\ includes a
degree of erosion in the use of B: the smaller the value of }$x^{\ast }$, 
\textit{the higher the matching probability }$(1-x^{\ast })(1-x^{\ast })$%
\textit{\ of two bilinguals who have shifted to speak A.\ }

According to this Proposition, the observed low levels of language $B$ use
could be theoretically explained by a convention with a relatively high
proportion of bilinguals choosing $H$ in equilibrium, $1-x^{\ast }$. It may
happen that, for values of $x^{\ast }\in (0,0.293)$, the matching
probability between bilinguals who will use $A$\ will be higher than the
matching probability of bilinguals who will use $B$: $(1-x^{\ast
})(1-x^{\ast })>x^{\ast }\times x^{\ast }+x^{\ast }(1-x^{\ast })+(1-x^{\ast
})x^{\ast }$. As any other convention (Young, 1995), the linguistic one,
will become a self-enforcing mechanism of language coordination, hard to
remove, which could extend to domains where information is not private, such
as family, friends, and assortative matchings\footnote{%
The following comments of the Secretary of Linguistic Policy for Basque,
after the \textit{V Sociolinguistic Survey,} are relevant: \textquotedblleft
In the last twenty years the use of Basque in the family has decreased one
per cent.\textquotedblright\ Baztarrika (2014). For the \textit{Surveys} of
Basque, see Appendix 6.1.}. The strong stability properties of $x^{\ast }$
allow us to make the following hypothesis to understand the use of $B$
reflected in the data:

\begin{proposition}
\label{Hypothesis} For a given proportion of bilinguals $\alpha \in (0,1)$,
and payoffs $m$, $n$ and $c$, the evolutionary stable mixed strategy Nash
equilibrium $x^{\ast }$could be thought of as a theoretical representation
(that is, a model) of the fraction of bilingual speakers who, in
real-life-situations, use language $B$ whenever they meet a bilingual.
\end{proposition}

We have data about the precise proportions of those who have the knowledge
of Basque, Irish, and Welsh, the $\alpha $ proportions. We also have
observations of the measures of actual use of $B$. To test empirically the
validity of this hypothesis, we build the function $x^{\ast }(\alpha )$ that
tells us the evolutionary stable Nash equilibrium associated to each $\alpha
\in (0,1)$.\footnote{%
You may consider $\alpha $ as an exogenous variable because its changes are
mainly due to factors related to cultural identity, and the responsibility
in transmitting the language to future generations. The existence of a
public bilingual educational system produces a continuous flow of bilinguals
that changes $\alpha $. This means that the Selten and Pool (1991)'s
argument of communicative benefits as the driving force to invest in the
learning of a language hardly applies here.} Based on $x^{\ast }(\alpha )$,
we, then build empirical models which will predict the \textit{street and
daily use} of $B$ of equilibrium. We test the validity of the hypothesis by
studying how good those predictions fit the observed data.\footnote{%
Note that $x^{\ast }(\alpha )$ and the predicting functions of language $B$%
-use, called $PKE(\alpha )$ and $PDU(\alpha )$ below, are not the same. You
can only identify and estimate $PKE(\alpha )$ and $PDU(\alpha )$, not $%
x^{\ast }(\alpha )$.}

\section{A Behavioural Theory for the Nash Equilibrium Function}

\label{sec-Nashequ}

To build the Nash equilibrium function $x^{\ast }(\alpha )$ we need to
develop a theory which would convert (\ref{eqn(1)}) into an econometric
model from which one can estimate empirical predictive models. But notice
that $x^{\ast }$ is obtained for a given $\alpha $ and constant $m$, $n$ and 
$c$. One can easily argue and empirically show that this is too strict for
being considered a realistic model. To prove this, we have made elsewhere
the exercise of predicting the use of $B$ based on the model of $x^{\ast }$
with $m$, $n$ and $c$ constant. We show that the model adapts badly to all
observed data, and was rejected throughout by nonparametric tests.
Furthermore, for $m$, $n$ and $c$ constant, $x^{\ast }(\alpha )$ is a
strictly increasing and concave function that can hardly be justified by
economic theory. So we need to take care of the parameters $m$, $n$ and $c$.%
\footnote{%
Since there are no data that allow to test some of the arguments we will use
in the following, one can only test the finally resulting model. This is a
problem shared by probably all papers testing economic theory with data. The
best we can do here is to develop a theory based on stylized facts, and
fundamental concepts of behavioural economics.}

Notice that one can think of $n$ as the natural pay-off level get by
monolinguals, and by the bilinguals who voluntarily choose language $A$.
Therefore, we may assume that $n$ is not sensitive to $\alpha $, and limit
to the modelling of $m$ and $c$.

The theory for $x^{\ast }(\alpha )$ is based on Kahneman and Tversky
(1979)'s psychological reference point, which is a cornerstone in
behavioural economics. Bilinguals may have different views and expectations
about factors that might affect the status of the minority language. They
would assign different probabilities to the event of, say, language shift,
they may also have different views about when $B$ is out of danger, or when
a {bilingual society} is achieved.\footnote{%
By bilingual society we mean a society where almost all individuals speak or
at least understand $A$ and $B$, and the obstacles to use $B$ are removed.}
They may also differ on how much their \textit{experienced utility}\footnote{%
That is, the hedonic experience associated with speaking $B$ or the painful
experience of being forced to use $A$. Henceforth, we will use indistinctly
`experienced utility' and `utility'.} (see Kahneman et al., 1997) is
increased when coordinating in $B$, that is, $m$, and on the loss of utility
when forced to switch to $A$, that is, $c$. Our view is that all the
differences in perceptions and expectations bilingual individuals might
have, are conditioned by the differences in the sociolinguistic contexts in
which they are immersed. By sociolinguistic context we refer to an area with
a specific proportion of bilinguals, and thus a certain presence of language 
$B$ in the workplace, streets, etc. (each context has a linguistic
landscape; see Landry and Bourhis, 1997) which clearly depends on $\alpha $.
For instance, the sociolinguistic context of the heavily industrialized
metropolitan area of Bilbao (23\% of Basque speakers and a $KE$ of 2.5\% )
is quite different from that of, say, San Sebastian-Donostia (with 40.6\%
and 15.2\%, respectively). The same applies to the context of Dublin city
and suburbs (32.8.2\% of Irish speakers, and a $DU$ of 1.3\%) relative to
the towns and villages of the Gaeltacht areas, such as Galway and suburbs
(41.4\%, and 3\%, respectively). Similarly, Cardiff (11.0\% and 4.86\%,
respectively) relative to Gwynedd (66.75\% and 66.75\%, respectively). It
would be highly unrealistic to assume that $m$ and $c$, are not sensitive to
the sociolinguistic contexts, i.e.\ to changes in $\alpha $; or,
equivalently, that they are the same in different municipalities and areas
with different $\alpha $.

Our theory assumes that each bilingual individual is adapted to some
sociolinguistic context of $B$ in relation to $A$, in the context of a
globalised society. The intuition is based on a basic tenet of Kahneman and
Tversky (1979): 
\textquotedblleft Our perceptual apparatus is attuned to the evaluation of
changes or differences rather than to the evaluation of absolute magnitudes.
When we respond to attributes such as brightness, loudness, or temperature,
the past and present context of experience defines an adaptation level, or
reference point, and stimuli are perceived in relation to reference point
[...]. Thus, an object at a given temperature may be experienced as hot or
cold to the touch depending on the temperature to which one has adapted. The
same principle applies to non-sensory attributes such as health, prestige,
and wealth.\textquotedblright\ 

We apply this idea to the perception of bilinguals about the minority
language $B$. In a language contact context, the frequency with which a
bilingual experiences meeting another bilingual in the past determines an
adaptation level or reference point. This adaptation conditions the
bilingual's expectations of the same event in the future, and of the status
of $B$. It will condition too his \textit{experienced utility} of speaking $%
B $. Each sociolinguistic context will shape a specific reference point for
him. Thus, the bilinguals' similarity of behaviour is sociolinguistic
context dependent.

Where $\alpha $ is high, the event of two bilinguals matching and speaking
in $B$ is fairly frequent. In those contexts bilinguals will be adapted to
that frequency. If $B$ is the dominant language of communication of a
sociolinguistic context, the utility bilinguals experience cannot be far
from the payoff $n$, i.e.\ bilinguals' utility $m$ would be near $n$.
Equivalently, in that sociolinguistic contexts the bilingual would not
suffer a \textit{perceptible utility loss} due to being forced to use $A$
when meeting, eventually, a monolingual. That is, we would assume that then $%
c$ would be `almost' zero.

In general, speakers of a minority language in a contact situation,
satisfying \textbf{A.1}, tend to feel a certain level of \textit{latent
frustration} caused by the permanent difficulties of keeping the language
alive. When the proportion of bilinguals is low, the probability of matching
between bilinguals is low, and so the probability of speaking in $B$ could
even be smaller. The bilinguals of that kind of context would be adapted to
those frequencies. The lack of use of $B$, and the overwhelming presence of $%
A$, induce some individuals to loose language skills and competence in $B$.
But others will try to escape by building ways to meet other bilinguals and
satisfy their wish of speaking $B$. So they make the effort of creating
clubs of $B$ speakers, spend personal time and effort for organising
activities in their leisure time (music gigs, drama productions, talks,
sports, language teaching for adults, and other events), just for the
pleasure of speaking $B$, and feeling that they are part of a cultural
identity\footnote{%
These types of clubs abound in the countries under study. In the Basque
Country they take the name of \textit{Mintzalagunak} \textquotedblleft Speak
with Friends\textquotedblright ; in Ireland are called \textit{Na Gaeil \'{O}%
ga GAA Club}; and in Wales \textit{Mentrau Iaith Cymru}.}. Clearly, the
experienced utility felt by these bilinguals when they speak $B$ must be
high.

If we take into account different municipalities or areas of the countries
or regions under study, we get a rich set of values of $\alpha $,
representing the variety of sociolinguistic contexts for $B$; that is, we
assume $\alpha \in (0,1)$.\footnote{%
We take the open unit interval $(0,1)$ as representing all the variety of
sociolinguistic contexts of the country.} We summarize the above as follows:
Let $\alpha _{L}^{\ast }\in (0,1)$ denote a hypothetical proportion of
bilinguals which the present bilinguals of the sociolinguistic context L
think it would convert L into a bilingual society\footnote{%
That is, a place, where there would be `almost' no obstacles to the social
use of $B$. It should not be a forecast based on sophisticated techniques.
It could only be a projection of desires to overcome the difficulties faced
by language $B$ in the present.}. Thus, $\alpha _{L}^{\ast }$ represents the
aspiration for the bilinguals of L\footnote{%
In a choice problem, the aspiration represents the most desired alternative,
available or not. Hence, at $\alpha _{L}^{\ast }$ the bilinguals, with
context L's preferences, feel satisfied.}.

\textbf{A.4.} \textit{Let }$S(\alpha _{L})=\alpha _{L}^{\ast }$\textit{\ be
the function that describes how the bilinguals of context }$L$\textit{\
determine their aspiration }$\alpha _{L}^{\ast }$\textit{, given the present 
}$\alpha _{L}$\textit{\ to which they are adapted to }$(1>\alpha _{L}^{\ast
}>\alpha _{L}>0)$\textit{. We assume that }$S$\textit{\ is a continuous,
concave, and strictly increasing function in }$\alpha $.

\textbf{Remark}: Since aspiration levels will differ due to the
heterogeneity of sociolinguistic contexts, we can assume that the distance
between the smallest aspiration level $\alpha _{\underline{L}}^{\ast }$ and
the highest one $\alpha _{\overline{L}}^{\ast }$ is small \footnote{%
That is, $\left\vert \alpha _{\overline{L}}^{\ast }-\alpha _{\underline{L}%
}^{\ast }\right\vert \leq \varepsilon $, for $\varepsilon >0$ small. This
would imply a extreme concave curvature of the $S$ function,}. Then we
choose the highest one to represent the aspiration level of the whole
bilingual population, and denote it as $\alpha ^{\ast }<1$. It may happen
that for a small community of bilinguals of some area $L$, $\alpha ^{\ast }$
is infeasible or out of reach, due to the cultural and sociological
constraints of that area. It is well known experimentally and theoretically,
that unavailable choices have an important influence on the decision
behaviour of agents, known as {aspiration effects} (Begum et al. 2016, and
Begum and Richter, 2015). The higher the distance between the actual $\alpha
_{L}$ and $\alpha ^{\ast }$, the higher the aspiration effect. Then $B$
would be perceived in a more intense way in contexts with low presence of
bilinguals, and some of them would volunteer to do extra work to save the
language in this area (as observed).

\textbf{A.5.} \textit{The experienced utility of speaking language }$B$%
\textit{, is a real valued continuous function, denoted }$m(\alpha )$\textit{%
, that strictly decreases with }$\alpha $\textit{. As }$\alpha $\textit{\
tends to the bilinguals' aspiration }$\alpha ^{\ast }$\textit{, }$m(\alpha )$%
\textit{\ tends to the "natural" pay-off level }$n$\textit{.}\footnote{%
Most likely, there would be a small interval of low levels of $\alpha >0$,
in which $m$ will be increasing in that interval. But for the greatest part
in the domain of $\alpha \in (0,1)$, $m(\alpha )$ is monotonically
decreasing. Nothing relevant is lost by keeping A.5 as it is.}

The \textit{latent} frustration mentioned above becomes \textit{explicit}
when the bilingual meets a monolingual. When this occurs, the level of
frustration felt will depend on the proportion $\alpha $, which is the
reference proportion or adaptation level for that bilingual. We denote the
(explicit) frustration felt in that matching as $c(\alpha )$. The reasons
given above lead us to make the following assumption.

\textbf{A.6}. \textit{The frustration cost of being forced to use }$A$%
\textit{\ is a real valued continuous function, denoted }$c(\alpha )$\textit{%
, that strictly decreases in }$\alpha $\textit{.}

\section{The econometric models}

\label{sec-empiric}

The next step is to apply \textbf{A.1}-\textbf{A.6} for building an
econometric model which we can be used for testing our model along the
observed data. We first specify $x^{\ast }$ to construct predictors for the
expected $KE$ and $DU$ along our theory model.

\subsection{From Nash equilibrium function to language use models}

We complete the building of $x^{\ast }$ as a function of $\alpha $ by
specifying $m$ and $c$ as simple functions satisfying \textbf{A.4-A.6}. Let
us suppose that $m(\alpha )$ is a simple decreasing function of type $\frac{K%
}{\alpha }$, where $K>0$ is a constant. For any $\alpha <\alpha ^{\ast }$ we
get $m(\alpha )>n$ such that a bilingual speaker get a positive profit
whenever he coordinates in $B$. Profits would be zero when the language is
normalized, in the sense $m(\alpha ^{\ast })=\frac{K}{\alpha ^{\ast }}=n$.
Then for $\alpha <\alpha ^{\ast }<1$, and a given value of $n$, we get $%
K=\alpha ^{\ast }n<n$ , and so the weighted profit $b(\alpha )=(m(\alpha )-n)%
\frac{\alpha }{(1-\alpha )}$ is a decreasing function.

Assumption \textbf{A.3} reads now as follows: for any $\alpha <\alpha ^{\ast
}$, the frustration cost function is strictly smaller than the weighted
benefit function$:$ 
\begin{equation*}
c(\alpha )<(m(\alpha )-n)\frac{\alpha }{1-\alpha }
\end{equation*}%
This condition allows the use of language $B$ in equilibrium (IU). Then 
\begin{equation}
c(\alpha )=(m(\alpha )-n)\frac{\alpha }{(1-\alpha )}-\tilde{b}(\alpha )
\label{eqn(2)}
\end{equation}%
for some $\tilde{b}(\alpha )>0$ denoting the net benefit. Inserting equation
(\ref{eqn(2)}) in (\ref{eqn(1)}) we obtain 
\begin{equation*}
x^{\ast }(\alpha )=\frac{\alpha (m(\alpha )-n)-c(\alpha )(1-\alpha )}{\alpha
(m(\alpha )-n)}=\frac{\tilde{b}(\alpha )(1-\alpha )}{\alpha (m(\alpha )-n)}\
,
\end{equation*}%
and substituting $m(\alpha )=\frac{K}{\alpha }$ gives 
\begin{equation}
x^{\ast }(\alpha )=\frac{\tilde{b}(\alpha )(1-\alpha )}{K-n\alpha }\ .
\label{eqn(3)}
\end{equation}%
Equation (\ref{eqn(3)}) shows how the predicted equilibrium proportion of
the bilingual population playing $R$ changes with $\alpha $. The denominator
is positive because $m(\alpha )=\frac{K}{\alpha }>n$ for all $0<\alpha
<\alpha ^{\ast }$. Hence $x^{\ast }=x^{\ast }(\alpha )>0$, and it is
increasing in $\alpha $ if $\tilde{b}(\alpha )$ is not decreasing faster
than $\frac{1-\alpha }{\alpha (m(\alpha )-n)}$ is increasing.\footnote{%
The first derivative of $g(\alpha )=\frac{1-\alpha }{\alpha (m(\alpha )-n)}$
with respect to $\alpha $ is $g^{,}(\alpha )=\frac{n-K}{(K-n\alpha )^{2}}$.
Hence, for $\alpha <\alpha ^{\ast }<1$, $g^{,}(\alpha )>0$ since $K=\alpha
^{\ast }n<n$.} So one would assume that the net benefit function $\tilde{b}%
(\alpha )$ has a well defined first derivative. Finally, let $\check{b}:=%
\tilde{b}/n$ denote the relative net benefit (recall, $n$ is constant).
Substituting $K=n\alpha ^{\ast }$ in the denominator of (\ref{eqn(3)}), we
get 
\begin{equation}
x^{\ast }(\alpha )=\frac{\check{b}(\alpha )(1-\alpha )}{(\alpha ^{\ast
}-\alpha )}\ .  \label{eqn(4)}
\end{equation}%
Since $x^{\ast }(\alpha )$ is an interior equilibrium, that is, $x^{\ast
}(\alpha )\in (0,1)$, we have $\check{b}(\alpha )(1-\alpha )<(\alpha ^{\ast
}-\alpha )$, where $1>\alpha ^{\ast }>\alpha >0$, and hence $0<\check{b}%
(\alpha )<1$.

We capture \textbf{A.6} with the specification $\check{b}(\alpha )=\beta
_{1}(\alpha ^{\ast }-\alpha )^{\beta _{2}}$ for unknown $\beta _{1}$, $\beta
_{2}$. We specify $\alpha ^{\ast }=\alpha ^{\beta _{3}}$ for unknown $\beta
_{3}$. For the model to make sense, we work with restrictions $\beta _{1}>0$
and $0<\beta _{3}<1$. Then (\ref{eqn(4)}) is converted into 
\begin{equation}
x^{\ast }(\alpha )=\ \beta _{1}(1-\alpha )(\alpha ^{\beta _{3}}-\alpha
)^{\beta _{2}-1}\ .  \label{(eqn(5))}
\end{equation}

As mentioned, our data of use, $KE$ and $DU$, refer to different things:
street use and daily use, respectively. This requires different modelling
for $KE$ and $DU$, the former being more challenging, as we will see.

\textbf{Model 1}: Having in mind (\ref{eqn(1)}), the model of expected $KE$
is obtained by looking at the probabilities of random matches of bilinguals
(if a monolingual is involved, $A$ is used anyhow) in which an $R$ player
does participate. That is, the probability of a random match where an $R$
player meets another $R$ player ($\alpha ^{2}{\ x^{\ast }}^{2}$), plus the
probability that an $R$ addresses an $H$ player ($\alpha ^{2}{x^{\ast }}%
(1-x^{\ast })$), plus the probability that an $H$ addresses an $R$ player ($%
\alpha ^{2}(1-x^{\ast }){x^{\ast }}$). In sum, 
\begin{equation}
E[KE|\alpha ,x^{\ast }]=\alpha ^{2}\left( 2x^{\ast }-{x^{\ast }}^{2}\right)
\ .  \label{model-a-KE}
\end{equation}%
This is the result of an $R$ player who takes the role of leader (as a
speaker or as a responder) in the language $B$ coordination process. When he
meets a bilingual, then they speak in $B$. An $R$ player will lead the
coordination process in any player role or position (row or column) in the
game. Let us call him strong $R$-player.

In real life however, we find bilinguals who sometimes lead the language $B$
coordination process, and sometimes are led. The reason is because
bilinguals are aware that, in order to satisfy their preference for the use
of $B$ requires being always the leader in the coordination. But that role
is very demanding. As mentioned, $R$ is a risky choice since you may suffer
a frustration if you meet a monolingual, and thus, get the minimum payoff, $%
n-c$. Further you may upset the monolingual and introduce tension in the
dialog alignment. It is natural then that a bilingual will try to avoid
being seen as rude or impolite. We should also have in mind that any
bilingual has a prior information about the rates of failure and success in
the coordination in $B$. That personal experience will shape his actual
behaviour, which moves him to be more flexible, combining the $R$ and $H$
strategies. Thus, in practise, the following model might be more realistic.

\textbf{Model 2}: Without loss of generality, we assume that in 50\% of all
his conversations the bilingual starts the conversation. He certainly does
so in $B$. The other 50\% are started by the other person, and he answers in
whatever language is used by the first speaker. We may call this bilingual a
weak $R$-player.

The resulting expected $KE$ is the sum of the probability of a random match
where an $R$ player meets another $R$-player ($\alpha ^{2}{\ x^{\ast }}^{2}$%
), plus the probability that an $R$-player addresses an $H$ player ($\alpha
^{2}{x^{\ast }}(1-x^{\ast })$), i.e. 
\begin{equation}
E[KE|\alpha ,x^{\ast }]=\alpha ^{2}{x^{\ast }}^{2}+\alpha ^{2}x^{\ast
}(1-x^{\ast })=\alpha ^{2}x^{\ast }\ .  \label{model-b-KE}
\end{equation}

\textbf{Model 3}: For modelling the expected daily use, $DU$, it is assumed
that almost all individuals that play strategy $R$ will answer (in the
census or survey) that they use $B$ every day, whereas almost all
individuals playing strategy $H$ will answer that they do not. Deviations
from these rule should cancel out in average. Consequently, the expected $DU$
is simply 
\begin{equation}  \label{model-DU}
E[DU|\alpha ,x^{\ast }]=\alpha x^{\ast } \ .
\end{equation}%
We propose the following theory models:

\begin{proposition}
\label{hypotheticModel} Let $\alpha $ denote the proportion of bilingual
speakers in a certain sociolinguistic context. Then

Model 1: $PKE_{1}(\alpha )=\alpha ^{2}(2x^{\ast }-{x^{\ast }}^{2})$ is the
predicted street use of $B$ in that context if the choosers of the $R$
strategy are strong.

Model 2: $PKE_{2}(\alpha )=\alpha ^{2}{x^{\ast }}$ is the predicted street
use of $B$ if the choosers of $R$ are weak.

Model 3: $PDU(\alpha )=\alpha {x^{\ast }}$ is the predicted daily use.
\end{proposition}

Note that, as already indicated, the data allow us to test our hypothesis
(say, Propositions \ref{Hypothesis} and \ref{hypotheticModel}) only as a
whole, not separately.

Using (5), we substitute $x^{\ast }$ in the three models, and get: 
\begin{eqnarray}
PKE_{1}(\alpha ) &=&\alpha ^{2}\left\{ 2\beta _{1}(1-\alpha )(\alpha ^{\beta
_{3}}-\alpha )^{\beta _{2}-1}-\left( \beta _{1}(1-\alpha )(\alpha ^{\beta
_{3}}-\alpha )^{\beta _{2}-1}\right) ^{2}\right\} \ ,  \label{PKE1} \\
PKE_{2}(\alpha ) &=&\alpha ^{2}\beta _{1}(1-\alpha )(\alpha ^{\beta
_{3}}-\alpha )^{\beta _{2}-1}\ ,  \label{PKE2} \\
PDU(\alpha ) &=&\alpha \beta _{1}(1-\alpha )(\alpha ^{\beta _{3}}-\alpha
)^{\beta _{2}-1}\ .  \label{PDU}
\end{eqnarray}%
These are the models we will study empirically for Basque, Irish and Welsh.
In order to estimate them from the samples $\{KE_{cti},\alpha
_{cti}\}_{i=1}^{n_{ct}}$, $\{DU_{cti},\alpha _{cti}\}_{i=1}^{n_{ct}}$ for
language sample $c$ in year $t$ one might consider either 
\begin{equation}
KE_{cti}=PKE_{cti}+\varepsilon _{cti}\ ,\quad DU_{cti}=PDU_{cti}+\epsilon
_{cti}\ ,  \label{eqn(5)}
\end{equation}%
(and analogously for $DU$) or, as a model with multiplicative structure, 
\begin{equation}
\log (KE_{cti})=\log (PKE_{cti})+\tilde{\varepsilon}_{cti}\ ,\quad \log
(DU_{cti})=\log (PDU_{cti})+\tilde{\epsilon}_{cti}\ .  \label{log-model}
\end{equation}%
This is done by least squares under the constraints that $\beta _{1}>0$ and $%
0<\beta _{3}<1$. Note that models (\ref{eqn(5)}) and (\ref{log-model})
require different assumptions on the (random) deviations from the mean. They
are, however, not testable, so that we have no particular a priori
preference. While the general findings are quite similar resulting from one
or the other estimation strategy, predicting the language use from the
logarithmic version (and consequently $\widehat{\log (KE)}$ or $\widehat{%
\log (DU)}$) is somewhat more complex as one has to correct for the - in our
case heteroscedastic - error dispersion since $E[\log (KE)|\alpha ]<\log
E[KE|\alpha ]$; the same holds certainly for $DU$. Also, the data fits when
looking at the untransformed data did not look very convincing. We therefore
give the least square estimates resulting from model (\ref{eqn(5)}). All
estimates are given together with nonparametric fits of $KE$ ($DU$) on $%
\alpha $ using local linear estimators and local bandwidths\footnote{%
Specifically, we used the command \texttt{locfit} of the R-package 'locfit'
with degree $1$ and nearest neighbour fraction set to 0.25 for Basque and
Welsh, but the 0.7 for Irish due to the extremely asymmetric distribution.
These settings were also used for the test statistics.} with Epanechnikov
kernel; see Appendix for details.

\subsection{Empirical Results}

\label{subsec-estimates}

\begin{figure}[tbh]
\begin{center}
\includegraphics[width=15.0cm,height=10cm,angle=0]{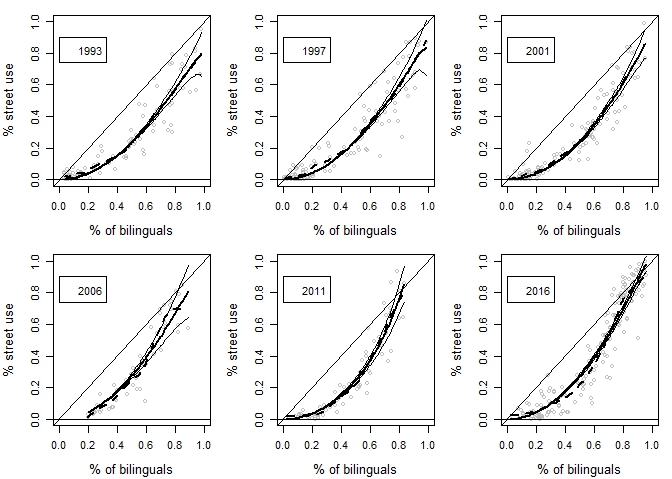}
\end{center}
\caption{Solid lines: $PKE$ for Basque, thick line is the estimate, thin
lines give the 99\% confidence bands. The dashed lines are nonparametric
kernel regression fits.}
\label{fig8}
\end{figure}

\begin{figure}[tbh]
\begin{center}
\includegraphics[width=15.0cm,height=5.0cm,angle=0]{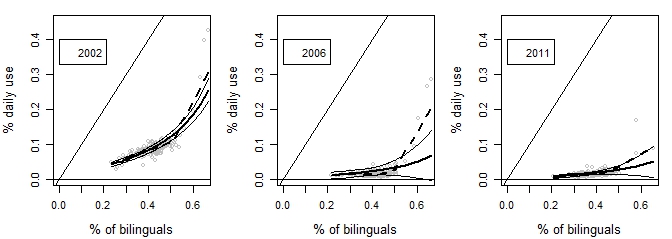}
\end{center}
\caption{Solid lines: $PDU$ for Irish, thick line is the estimate, thin
lines give the 99\% confidence bands. The dashed lines are nonparametric
kernel regression fits.}
\label{fig9}
\end{figure}

\begin{figure}[tbh]
\begin{center}
\includegraphics[width=10.cm,height=5.0cm,angle=0]{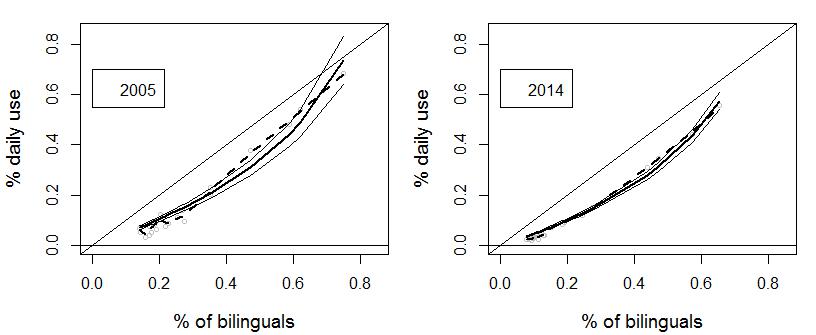}
\end{center}
\caption{Solid lines: $PDU$ for Welsh, thick line is the estimate, tin lines
give the 99\% confidence bands. The dashed lines are nonparametric kernel
regression fits.}
\label{fig10}
\end{figure}

The presentation of results concentrates on the predictive functions, as
these are in the centre of our interest. The parameters have no interesting
interpretation, and are therefore deferred to the Appendix 6.3 and 6.4.
While the 99\% confidence bands of the functions are very narrow, the 95\%
confidence intervals for the parameter estimates are quite wide due to
extremely high correlations (not shown) between the parameter estimates.
These, like standard errors and the p-values of our test, are estimated by
wild bootstrap. In Figures \ref{fig8} to \ref{fig10} the thick solid line is
the model based estimate with confidence bands indicated by thin solid
lines. The dashed lines are the nonparametric fits, and the grey circles
indicate the recorded observations. To see better the curvature, we have
also plotted a $45^{o}$ line, and a horizontal line at $\alpha =0$. The
functions are estimated from different samples taken from the sources
discussed in Section \ref{sec-KEdef}. For Basque we have data from between
53 to 175 municipalities per year (from \textit{Soziolinguistika Klusterra}%
), for Irish from about 180 so-called 'local electoral areas' obtained from
different censi, and for Welsh from 22 local authorities, cf.\ Tables \ref%
{tab1} to \ref{tab3} in the Appendix.


For both, Basque and Irish we see that, over time, the functions are quite
stable, even over a period of almost 25 years.\footnote{%
For Ireland we mainly refer to 2006 and 2011 because of the uncertainty
concerning the measurement of $DU$ before 2006. People may use every day
Irish only at school, but elsewhere choose strategy $H$. Only since 2006
there is a clear definition of $DU$ outside the educational system.} The $PDU
$ for Welsh is almost parallel to the $45^{o}$ line. When $\alpha $ is above 
$0.6$, i.e.\ bilinguals are a clear majority, then the function immediately
shifts to the $45^{o}$ line. All model-based estimates come very close to
the nonparametric (i.e.\ model-free) data fits, except maybe for extreme
values and outliers. In particular, most of the time the confidence bands
include the nonparametric fit. All this clearly confirms our theory
discussed in the previous sections, and thus, supports our \textit{extended}
LUG model.

Further, for Irish we see a clear difference compared to Welsh and Basque.
The latter two have a more intensive use than Irish, which seems to be
shifted away. For Basque, it even seems that the curvature is getting closer
to the $45^{o}$ line; this is studied further below. But before doing so, we
complete the empirical study of our hypothesis, Propositions \ref{Hypothesis}
and \ref{hypotheticModel}, by another formal test. We say `another' since
the provision of confidence bands is equivalent to a particular statistical
test.

More specifically, we apply the nonparametric specification test of H\"{a}%
rdle and Mammen (1993) which takes as test statistic the Euclidean distance
between the smoothed version of the model-based parametric fit and the
nonparametric fit. The `smoothed version' means that the parametric fit gets
convoluted with the same kernel and bandwidth we used for the nonparametric
fit to avoid rejecting our hypothesis simply because of a smoothing bias
inherited by the nonparametric fit. Then the bootstrap p-values are: for
Basque $0.728$ for 1993, $0.408$ for 1997, $0.559$ for 2001, $0.054$ for
2006, $0.479$ for 2011, and $0.003$ for 2016. For Irish $0.055$ for 2002, $%
0.032$ for 2006, and $0.095$ for 2011; and for Welsh we get a p-value of $%
0.075$ for 2005 and $0.080$ for 2014. This, together with the graphs,
indicate that the model adapts reasonably well to the observed data. The
reason for the low p-values in some years, that can also be detected in the
pictures, is due to the $\alpha $ values for which the nonparametric line is
far outside the confidence bands.

To calculate the p-values, uniform confidence bands, standard errors and
confidence intervals, we generated 1000 bootstrap samples $\{Y_{i}^{\ast
},\alpha _{i}\}_{i=1}^{n}$ where $Y_{i}^{\ast }:=\widehat{Y_{i}}+(Y_{i}-%
\widehat{Y_{i}})\cdot \varepsilon _{i}$ with $Y_{i}$ being $KE_{i}$ (or $%
DU_{i}$), $\widehat{Y_{i}}$ our estimated prediction model, and $\varepsilon
_{i}$ randomly drawn from a standard normal\footnote{%
We also checked with centred and normalised chi-square to guarantee
non-negative responses (Pendakur, Scholz, Sperlich, 2010), and a two-point
distribution (H\"{a}rdle, Mammen, 1993) to account better for potential
asymmetries. We present the version with standard normal $\varepsilon _{i}$
as these seem to calibrate best nonparametric test for moderate sample
sizes, see Sperlich (2014).}. Then, for example the uniform confidence bands
for \textit{PKE} were constructed in the following way. Define 
\begin{equation*}
T:=\mathrel{\mathop{sup}\limits_{\alpha\in (0,1)}^{}}|PKE(\alpha )-\widehat{%
PKE}(\alpha )|/\sigma _{y}(\alpha )
\end{equation*}%
with $\sigma _{y}^{2}(a)=Var[\widehat{PKE}(a)]$. It can be shown that its
bootstrap analogue 
\begin{equation*}
T^{\ast }:=\mathrel{\mathop{sup}\limits_{\alpha\in (0,1)}^{}}|PKE^{\ast
}(\alpha )-\widehat{PKE}^{\ast }(\alpha )|/\sigma _{y}^{\ast }(\alpha )
\end{equation*}%
is converging in distribution to $T$. Consequently, from our bootstrap
samples we can obtain any quantile $q_{T}$ of $T$. \ The formula $\widehat{%
PKE}(\alpha )\pm q_{T}^{\ast }\sigma _{y}^{\ast }(\alpha )$ provides us with
uniform confidence bands at any wanted quantile. The confidence bands (as
the confidence intervals) can be asymmetric because they are calculated by
this bootstrap method, and therefore reflect the asymmetry of the
distributions of our estimators.

These uniform bands are very useful as the nonparametric test simply tells
us whether the theory model has a low p-value or not; it does not tell us
where a potential problem is. In our case, the bands provide us with the
information that there is not a general, systematic deviation of our theory
model from the nonparametric fit. In fact, apart from outlier problems on
the right hand side of the Irish data, the low p-values for Basque in 2016
and Irish in 2006 are due to an 'elbow' of the real data (respectively the
nonparmaetric fit) at around $\alpha =0.5$, i.e.\ when bilinguals become the
majority.

\begin{figure}[hbt]
\begin{center}
\includegraphics[width=13.0cm,height=5.0cm,angle=0]{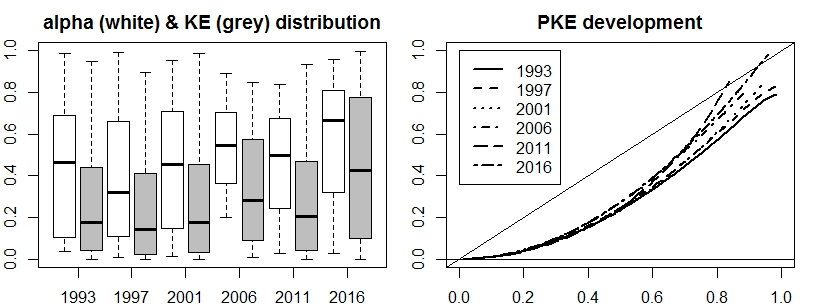}
\end{center}
\caption{The development of $\protect\alpha$ and street use of Basque over
time.}
\label{fig11}
\end{figure}
\begin{figure}[tbh]
\begin{center}
\includegraphics[width=13.0cm,height=5.0cm,angle=0]{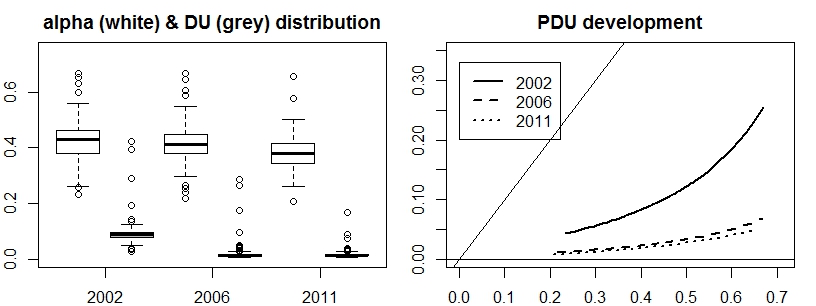}
\end{center}
\caption{The development of $\protect\alpha$ and daily use of Irish over
time.}
\label{fig12}
\end{figure}
\begin{figure}[tbh]
\begin{center}
\includegraphics[width=13.0cm,height=5.0cm,angle=0]{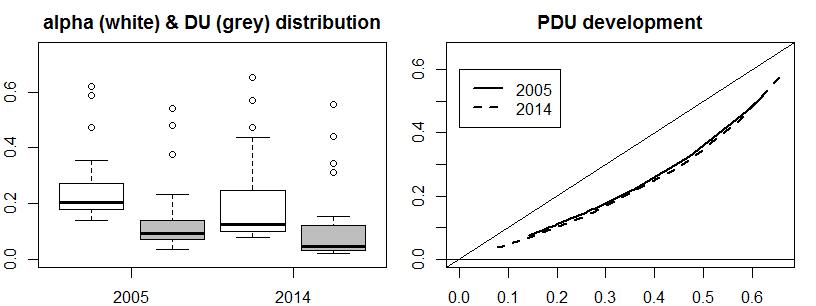}
\end{center}
\caption{The development of $\protect\alpha$ and daily use of Welsh over
time.}
\label{fig12}
\end{figure}

We said above that Irish is almost shifted, whereas the use of Basque is
quite stable. Having made these observations, it would be interesting to
contrast them with the development of  $\alpha $ and $KE$ (or $DU$) over
time. The box-plots (left panels) in Figures \ref{fig11}-\ref{fig12}
illustrate the development of the distributions over the years. Recall that
we are looking at all combinations $(\alpha _{ti},KE_{ti})$ without
weighting them by the population size of municipality $i$. This explains why
it seems that the percentage(s) of bilingual speakers went down, though the
real total percentage has steadily increased, see Table \ref{tab1Basque}. We
see that all years exhibit a huge dispersion for $\alpha $ and $KE$ (less so 
$DU$), with no stabilization of the distributions of these indicators. We
observe a shrinking number of municipalities with small $\alpha $ and/or
small values of $KE$. This might explain why people feel that $\alpha $ has
to increase much more to become a normalized language ($\beta _{3}$ is
converging to zero from above!). What we see in the right panels is the
development of $PKE$ and $PDU$ over time. The tendency is clear: for Irish
it went down dramatically from an almost low level, whereas for Basque it is
steadily increasing. As for Welsh we only have two points in time which
certainly give different results; thus, we prefer not to interpret the
observed change.

We close this empirical analysis by noting that we also run estimations and
test for Model 1, and for $x^{\ast }$ with $m$ and $c$ constant, as well as
with other flexible net benefit functions allowing to violate one of our
assumptions. They neither fit better but get smaller p-values. All these
results are provided on request. We can conclude that we could not find any
empirical evidence against our theory based model.

\section{Conclusions}

\label{sec-concl}

This paper has taken an evolutionary game-theoretic approach to analyse the
observed erosion in the minority language use in modern multilingual
societies, like the Basque Country, Ireland and Wales (the benchmark
societies). Certainly, manifold factors could influence a minority culture,
such as the developments of information technologies, globalization and
international market integration. Olivier et al.(2008), Bisin and Verdier
(2010) and (2014) study the impact of international integration on cultural
identities with the cultural dynamics model of Bisin and Verdier (2001)
interacting with a model international trade equilibrium. We do not deny the
influence of globalization on cultural identities. However, on a more
specific level, such as the analysis of minority language use, we rather
take an internal view, focusing, as suggested by \textit{the Council of
Europe,} on the speaker of the minority language $B$, i.e., the bilingual
individual, and the factors that might shape his language choices.

Our model shows that the bilingual population dynamics converge into an
(interior) evolutionary stable Nash equilibrium (i.e., a linguistic
convention), $x^{\ast }$, that denotes the equilibrium fraction of
bilinguals who choose pure strategy $Reveal(R)$. In this equilibrium, when
pairs of bilinguals playing $Hide(H)$ are matched, with probability $%
(1-x^{\ast })(1-x^{\ast })$, they will shift to speak in the majority
language $A$. Since the strategies $R$ and $H$ have equal expected payoffs,
the linguistic convention is compatible with any value of $x^{\ast }\in (0,1)
$. Then it is possible that $H$ could create a positive feedback loop; that
is, more bilinguals would choose it, increasing $1-x^{\ast }$. As a result,
a convention with a relatively low fraction of $R$ choosers may, by
imitation learning, expand through the bilingual population, reaching
domains where asymmetric information does not exist, such as family, and
assortative matchings\textit{.}

Hence, we have a theoretical explanation for the erosion observed in the
minority language use. To test the theory, we build three empirical
equilibrium models, each giving rise to an increasing and convex relation
between the proportion of bilinguals, $\alpha $, in each context and the
predicted use of $B$ in equilibrium\footnote{%
It could be argued that intuition might suggest that the predictive
functions should be increasing and convex. However, these two properties
alone cannot explain the variation observed in the data. There is, still,
plenty of space for model mis-specification.}. We show that these
equilibrium functions explain very well the observed data for all languages
and years. They are quite stable over time, for each language, and the
performed nonparametric tests do not reject our model. The empirical result
is that the \textit{extended} LUG is a solid model to analyse the erosion in
the use of language $B$.

One wonders why bilinguals could behave in that (paradoxical) way? The
strategic analysis of language use of this paper throws light to understand
this behaviour. It is well known that human language is guided by
economizing attitudes\footnote{%
Characterized by the optimization of the difference between communication
benefits and the costs attached to the speech production (in each of the two
active language system, in the case of a bilingual): memory, word length,
utterance length, frequency and other factors; see Frank and Goodman (2012).}%
, and the principles of least effort, see Zipf (1949).\footnote{%
These features are also confirmed by Selten and Warglien (2007) in their
experiment on the emerging of an artificial language.} Our framework points
in the same direction. More specifically, contrary to the strategy $R$, the
chooser of $H$ minimizes potential frictions with monolinguals, and takes a
passive role in determining the language of interaction. Further, $H$ avoids
the frustration cost $c$. Thus, it should not be a surprise that bilinguals
might converge into a linguistic convention in which the least effort
demanding strategy $H$ is chosen by a relevant fraction of the bilingual
population which, in turn, would result in a decrease in the use of $B$.

Since the goal is to increase the use of $B$, our model suggests that the
main policy target should be to increase $x^{\ast }(\alpha )$. Therefore,
all the elements entering into equation (3) should be taken into account: $%
\alpha (m(\alpha )-n)$, $c(\alpha )$ and $\alpha .$ Needless to say, the
imperfect information should be minimized. In fact, its influence on
minority language use has been neglected in language economics (see
Ginsburgh and Weber, Ed., 2016), sociolinguistics (Bayley, Cameron and
Lucas, Ed., 2013), and in the theories developed after the Abrams and
Strogatz (2003)'s model (see Patriarca et al., 2012). There are simple ways
to transmit information, like the Cymraeg badges proposed by the Welsh
Language Commission so that Welsh speakers would recognize each other. But
this kind of signalling could be controversial. For example, it would not be
politically accepted in the Basque Country. Indeed, the type of signalling
(verbal or non verbal) can only be adopted if it has a wide social
consensus. Verbal signalling could be more complicated. Linguistic planners
could consider the monolinguals as potential allies. To this end, careful
pondered introductory phrases could be designed, to be used by bilinguals to
convey the desire to speak $B$; also, well thought answers should be chosen
for the monolinguals to convey that they do not feel under pressure nor are
embarrassed to confess they do not speak the official language $B$. Since
the interactions with monolinguals are frequent, their cooperative attitude
would make bilinguals to gradually ignore politeness-based choices, and 
\textit{reveal} more openly their type. Friendliness could also alleviate
frustration costs.

It might require also language policy measures to increase the bilinguals'
perceived net benefit of using $B$. Policies could also highlight the
advantages (both cultural and psychological) of being an active bilingual,
taking part in keeping alive language $B$ and its related culture, and
sustain cultural diversity.

A dramatic increase in $\alpha $ is not necessarily the definitive solution
to increase the use of $B$, as it has been proved by comparing Ireland,
Wales, and the Basque Country: while the percentage of bilinguals in Ireland
doubles the one in the Basque Country, the former is close to extinction
while the latter exhibits a fairly stable street use. One may think that
this is because English (the majoritarian language $A$) is a stronger
competitor for Irish ($B$) than Castellano and French is for Basque (though
Spanish is after English, Chinese and Hindi the fourth most spoken language
in the world). For this reason we added Welsh which has an $\alpha $ similar
to the one of Basque.

Like any model, ours has limitations. There may be other factors outside the
model that hinder the use of $B$. For example, with the empirical Model 2 we
already capture partially the bilinguals' politeness behaviour, and find
that it explains better the data of street use of Basque (compared with
Model 1). This suggests that politeness is a relevant feature to understand
the use of $B$. One could extend the LUG further by introducing explicitly
politeness. This would imply a different theoretical scenario, with
different Nash equilibria and linguistic conventions; see Uriarte (2017) for
a preliminary study.

\section*{References}

\begin{description}
\item Abrams D.M. and Strogatz S.H. (2003). \textquotedblleft Modeling the
dynamics of language death,\textquotedblright\ \textit{Nature} 424, 900.


\item Altuna, O. and Basurto, A. (2013). \textit{A Guide to Language Use
Observation.} \textit{Survey Methods.} Soziolinguistika Klusterra.

\item Amorrortu, E.. Ortega, A. Idiazabal, I. and Barre\~{n}a, A. (2009). 
\textit{Actitudes y Prejuicios de los Castellanohablantes Hacia el Euskera}.
Eusko Jaurlaritza-Basque Government: Vitoria-Gasteiz.

\item Robert Bayley, Richard Cameron, and Ceil Lucas (eds.) (2013). \textit{%
The Oxford Handbook of Sociolinguistics}. Oxford. Oxford University Press.

\item Anderlini, L. and Felli, L. (2004). \textquotedblleft Book Reviews:
Economics and language,\textquotedblright\ \textit{Economica} 71, 169-179.

\item Baztarrika, P.(2014). \textquotedblleft Euskararen hazkundearen
paradoxak,\textquotedblright\ \newline
http://www.erabili.eus/zer\_berri/muinetik/.

\item Begum, G. and Richter, M. (2015). \textquotedblleft An experiment on
aspiration-based choice,\textquotedblright\ \textit{Journal of Economic
Behavior \& Organization} 119, 512--526.

\item Begum, G.,\ Richter, M. and Tsur, M. (2016). \textquotedblleft
Aspiration-based choice theory,\textquotedblright\ Working Paper.

\item Binmore, K. Gale, J. and Samuelson, L.(1995). \textquotedblleft
Learning to be imperfect: The ultimatum game,\textquotedblright\ \textit{%
Games and Economic Behavior} 8, 56-90

\item Bisin, A and Verdier, T. (2001). \textquotedblleft The economics of
cultural transmission and the dynamics of preferences,\textquotedblright\ 
\textit{Journal of Economic Theory} 97, 298-319.

\item \textsl{\ }Bisin, A., Verdier, T. (2010). \textquotedblleft The
economics of cultural transmission and socialization,\textquotedblright\ in
Benhabib, J., Bisin, A., Jackson, M. (eds.), \textit{The Handbook of Social
Economics}. Elsevier: New York.

\item Bisin, A and Verdier, T. (2014). \textquotedblleft Trade and Cultural
Diversity,\textquotedblright\ in Ginsburgh, V. and Throsby, D. (eds.) 
\textit{Handbook of the Economics of Art and Culture}, Vol.2. Amsterdam:
Elsevier.

\item Blume, A. and Board, O. (2013). \textquotedblleft Language
barriers,\textquotedblright\ \textit{Econometrica} 81(2), 781-812.

\item Brown, P. and Levinson, S.C. (1987). \textit{Politeness: Some
Universals in Language Usage}. Cambridge: Cambridge University Press.

\item Cabrales, A. (2000). \textquotedblleft Stochastic replicator
dynamics,\textquotedblright\ \textit{International Economic Review} 41,
451-481

\item Cabrales, A. and Uriarte, J.R. (2013). \textquotedblleft Doubts and
equilibria,\textquotedblright\ \textit{Journal of Evolutionary Economics}
23(4), 783-810.

\item Cavalli-Sforza, L. and Feldman, M. (1981).\textit{\ Cultural
Transmission and Evolution: A Quantitative Approach}. Princeton, NJ:
Princeton University Press.

\item \textit{Council of Europe}. http://www.coe.int/t/dg4/education/minlang.

\item Crystal, D. (2000). \textit{Language Death}, Cambridge: Cambridge
University Press.


\item Desmet, K., Ortu\~{n}o-Ort\'{\i}n, I. and Weber, S. (2009).
\textquotedblleft Linguistic diversity and
redistribution,\textquotedblright\ \textit{Journal of the European Economic
Association} 7(6), 1291-318.

\item Fishman, J.A. (1991). \textit{Reversing Language Shift. Theoretical
and Empirical Foundations of Assistance to Threatened Languages}. Clevedon,
UK: Multilingual Matters.

\item Fishman, J.A. (2001). \textquotedblleft Why is it so hard to save a
threatened language?\textquotedblright\ in J.A. Fishman (ed.) \textit{Can
Threatened Languages be Saved}? Clevedon, UK: Multilingual Matters.

\item Frank, M.C. and Goodman, N.D. (2012). \textquotedblleft Predicting
pragmatic reasoning in language games,\textquotedblright\ \textit{Science}
336(6084), 998.

\item Ginsburgh, V. and Weber, S. (2011). \textit{How Many Languages Do We
Need? The Economics of Linguistic Diversity}. Princeton, NJ: Princeton
University Press.

\item Ginsburgh, V. and Weber, S., Ed. (2016). \ \textit{The Palgrave
Handbook of Economics and Language}. Houndmills:\ Palgrave MacMillan.

\item Grin, F., Sfreddo, C. and Vaillancourt, F. (2010). \textit{The
Economics of the Multilingual Workplace}. New York, NY: Routledge.

\item H\"{a}rdle, W. and Mammen, E. (1993). \textquotedblleft Comparing
Nonparametric Versus Parametric Regression Fits,\textquotedblright\ \textit{%
Annals of Statistics} 21(4): 1926--1947.

\item \textit{International Journal of the Sociology of Language} (2015).
\textquotedblleft Special Issue: New Speakers of Minority Languages: The
Challenging Opportunity,\textquotedblright\ Volume 231.

\item Iriberri, N. and Uriarte, J.R. (2012). \textquotedblleft Minority
language and the stability of bilingual equilibria,\textquotedblright\ 
\textit{Rationality and Society} 2(4), 442-462.

\item Jones, H.M. (2012). \textit{A statistical Overview of the Welsh
language}. The Welsh Language Board. Cardiff.

\item Kahneman, D. and Tversky, A. (1979). \textquotedblleft Prospect
Theory: An analysis of decision under risk,\textquotedblright\ \textit{%
Econometrica} 47(2), 263-289.

\item Kahneman, D., Wakker, P.P. and Sarin, R. (1997). \textquotedblleft
Back to Bentham? Explorations of experienced utility,\textquotedblright\ 
\textit{The Quarterly Journal of Economics} 112(2): 375--405.

\item Landry, R. and Bourhis, R.Y. (1997). \textquotedblleft Linguistic
landscape and ethnolinguistic vitality an empirical
study,\textquotedblright\ \textit{Journal of Language and Social Psychology}
16 (1): 23--49.

\item Manski, Ch. (1997). \textquotedblleft Identification of anonymous
endogenous interactions,\textquotedblright\ in B. Arthur, S. Durlauf, and D.
Lane (eds) \textit{The economy as an Evolving Complex System II}, Reading,
Mass.: Addison-Wesley, 1997, pp. 369-384.

\item Maynard Smith, J. (1982). \textit{Evolution and the theory of games}.
Cambridge: Cambridge University Press.

\item Mercator (2008).\textit{\ The Basque language in Education in Spain}.
2nd Edition. \newline
www.mercator-research.eu

\item Mercator (2014). \textit{The Welsh language in Education in the UK}.
2nd Edition. www.mercator-research.eu

\item Mercator-Education (2016). \textit{The Irish language in education in
the Republic of Ireland}. 2nd Edition. www.mercator-research.eu

\item Minett, J. and Wang, W.-Y. (2008). \textquotedblleft Modelling
endangered languages: The effects of bilingualism and social
structure,\textquotedblright\ \textit{Lingua} 118(1), 19-45.

\item Nelde, P.H. (1995). \textquotedblleft Language contact and conflict:
the Belgian experience and the European Union,\textquotedblright\ in S.
Wright and H. Kelly (eds.) \textit{Languages in Contact and Conflict:
Experiences in the Netherlands and Belgium}. Clevedon, UK: Multilingual
Matters.

\item Olivier, J., Mathias T., and Verdier, T. (2008).\textquotedblleft
Globalization and the dynamics of cultural identity,\textquotedblright\ 
\textit{Journal of International Economics} 76, 356--370

\item Ortega, A., Urla, J. Amorrortu, E. Goirigolzarri, J. and Uranga, B.
(2015). \textquotedblleft Linguistic identity among new speakers of
Basque,\textquotedblright\ \textit{International Journal of the Sociology of
Language} 231, 85--105.

\item O'Rourke, B. and Ramallo, F. (2011). \textquotedblleft The
native-non-native dichotomy in minority language contexts. Comparisons
between Irish and Galician,\textquotedblright\ \textit{Language Problems and
Language Planning} 35(2): 139--159.,

\item Patriarca, M. and X. Castell\'{o}, Uriarte, J.R., Eguiluz, V.M. and
San Miguel, S. (2012). \textquotedblleft Modeling two Language competition
dynamics,\textquotedblright\ \textit{\ Advances in Complex Systems}, 15,3-4.

\item Pendakur, K., Scholz, M. and Sperlich, S. (2010). \textquotedblleft
Semiparametric indirect utility and consumer demand,\textquotedblright\ 
\textit{Computational Statistics and Data Analysis} 54: 2763--2775.

\item Rubinstein, A. (2000). \textit{Economics and Language, }Cambridge: CUP%
\textit{.}

\item Schlag, K. H. (1998). \textquotedblleft Why imitate, and if so, how?:
A boundedly rational approach to multi-armed bandits,\textquotedblright\ 
\textit{Journal of Economic Theory}, 78, 130-156.

\item Selleck, C. (2018). \textquotedblleft We are not fully Welsh.
Hierarchies of belonging and new speakers of Welsh,\textquotedblright\ in
Smith-Christmas, C., \'{O} Murchadha, N.P., Hornsby, M., Moriarty, M.
(eds.). \textit{New Speakers of Minority Languages: Linguistic Ideologies
and Practices}. Palgrave Macmillan, UK.

\item Selten R. and Pool, J. (1991). \textquotedblleft The Distribution of
Foreign Language Skills as a Game Equilibrium,\textquotedblright\ in R.
Selton (ed). \textit{Game Equilibrium Models IV: Social and Political
Interaction}, Berlin: Springer-Verlag, 64-87.

\item Selten, R. and Warglien, M. (2007). \textquotedblleft The emergence of
simple languages in an experimental coordination game,\textquotedblright\ 
\textit{Proceedings of the National Academy of Science} 104(18), 7361-7366.

\item Sol\'{e}, R., Corominas-Murtra, B. and Fortuny, J. (2010).
\textquotedblleft Diversity, competition, extinction: the ecophysics of
language change,\textquotedblright\ \textit{Interface} 7, 1647-1664.

\item Sperlich, S. (2014). \textquotedblleft On the choice of regularization
parameters in specification testing: a critical
discussion,\textquotedblright\ \textit{Empirical Economics} 47(2), 427--450.

\item Stauffer, D. and Schulze, C. (2005). \textquotedblleft Microscopic and
macroscopic simulation of competition between Languages,\textquotedblright\ 
\textit{Physics of Life Reviews} 2(2), 89-116.

\item \textit{UNESCO} (2002). "UNESCO Universal Declaration On Cultural
Diversity"

\item Uriarte, J.R. (2017). \textquotedblleft The Economics of Bilingual
identity: Politeness equilibrium,\textquotedblright\ Mimeo, University of
the Basque Country.

\item Weibull, J.W. (1995). \textit{Evolutionary Game Theory}. Cambridge,
Mass.: the MIT Press.

\item Winford, D. (2003). \textit{An Introduction to Contact Languages}.
Oxford: Blackwell.

\item Young, H.P. (1995). \textquotedblleft The economics of
convention.\textquotedblright\ \textit{Journal of Economic Perspectives} 10,
105-122.

\item Zipf, G. K. (1949). \textit{Human Behavior and the Principle of Least
Effort.} Cambridge, Mass: Addison-Wesley Press.
\end{description}

\section{Appendix}

\label{sec-appendix}

\subsection{Language Use Data}

Note that the proportion of bilinguals in the three languages, denoted $%
\alpha $, is obtained dividing the Number of Bilinguals by the Total
Population. For Table 1 this is the population in the country, and in the
Figures in Section 4 it refers to the small regions (communities, counties,
etc). The definition of who is bilingual is set by the statistical authority
in each country. They coincide in that it is the oral competence what
determines whether an individual is qualified as bilingual or not.

- \textbf{Basque Country}: The data about knowledge of Basque can be found
in the \textit{Sociolinguistic Surveys (Inkesta Soziolinguistikoa). }These
surveys are carried out by the local Government and agencies of each of the
three territories where Basque is spoken: the Basque Autonomous Community,
the Navarre Community and Pays Basque.\textit{\ }The population is divided
in three groups according to the oral competence relative to Basque:
Bilinguals (those who speak well or fairly well the language), Passive
Bilinguals (those who do not speak well, but understand the language) and
Monolinguals. The VI Sociolinguistic Survey 2016 gathered information from a
sample of randomly chosen 8200 people, aged 16 and over, who answered
questionnaire over the phone. See the Basque Government's gathered data
about Basque in: \texttt{%
www.euskara.euskadi.eus/r59-738/eu/contenidos/informacion/argitalpenak 
\newline
/eu\_6092/ikuspegi\_sozio\_linguis.html}

The \textit{Street Surveys} are carried out, every five years, by the NGO 
\textit{Soziolinguistika Klusterra.} The street surveyors collect random
observations, without any interference with the subjects observed, just by
listening and recording the language of conversations taking place in public
spaces of the above mentioned three Administrative authorities.\footnote{%
For the methodological details on how the data is collected, see Altuna and
Basurto (2013). Note that the \textit{V. Street Survey} in 2006 took place
in 62 municipalities with 400 street surveyors recording 185,316
conversations. In the \textit{VII. Street Survey} in 2016, 187,635
conversations in 144 municipalities were recorded.} Using random samples of
anonymously registered conversations in the streets at a given time and
place, the \textit{Street Use Measure} (\textit{KE}) of minority language $B$
is the number of individuals observed in conversations speaking language $B$
out of the total number of individuals observed in conversations in that
place. As a methodological rule, the observers only record conversations in
a single language, cf.\ \textbf{A.1}. The \textit{street use} data can be
found on \texttt{www.soziolinguistika.org}.

- \textbf{Ireland}: All data were obtained from the web-page of the Irish
statistical institute, http://www.cso.ie/en/census/index.html. However, they
clean up and change their web-pages regularly, such that the data we
downloaded for 1996, 2001/2 and 2005/6 cannot be found that easily anymore.
We certainly provide these data on request. The 2011 data are still
available but differently organized. Again we can provide them on request.
For the census 2016 of Ireland, so far only results and statistics are
available, but we did not find the disaggregated data that we needed for our
calculations. In all censi bilinguals are defined as "Irish speakers aged 3
years and over". The daily use, \textit{DU}, is explicitly observed, after
1996 separated into groups that daily use Irish "inside the educational
system only" and those who also or exclusively use it outside the
educational system. Certainly we consider the latter group. As for 1996 this
distinction is not at all clear we skipped this wave in the newer version of
our paper.

- \textbf{Wales}: The most reliable source of data about Welsh speakers are
the Censi of Wales. It is not explicitly used the word bilingual as they
consider \textquotedblleft the number of people aged 3 and over able to
speak Welsh\textquotedblright . Under the heading of \textquotedblleft able
to speak Welsh\textquotedblright\ is included every person aged 3 and over
who \textquotedblleft Speaks, reads and writes Welsh\textquotedblright\ or
\textquotedblleft Speaks, reads, but does not write\textquotedblright\ or
\textquotedblleft Speaks but does not read or write\textquotedblright . Data
about the frequency of use of Welsh are in the WLUS (\textit{Welsh Language
Use Surveys). }The daily use, \textit{DU,} is obtained dividing the number
of \textquotedblleft daily users\textquotedblright\ by the total population.
Disaggregated data for daily use of Welsh by local authorities is obtained
from Jones (2012). Used sources are:

\textit{The Welsh Language Use Surveys of 2004-06.} The Welsh Language
Board. 2008. Cardiff.

\textit{Welsh Language Use in Wales 2013-15.} Welsh Government and Welsh
Language Commissioner. 2015. Cardiff.

Jones, H.M. (2012). \textit{A Statistical Overview of the Welsh Language},
The Welsh Language Board. Cardiff.

\textit{2001 Census: Report on the Welsh Language.} Office for National
Statistics, London: TSO. 2004.

\textit{2011 Census. First results on Welsh}, Statistical Bulletin 118/2012.
Statistics for Wales.

\textbf{Additional data}: We send two \textit{Excel} \textit{files} with
data about Irish and Basque. In the latter case, the headings were written
in Basque with English acronyms by the surveyors who collected the data. The
headings in the columns are: Year (1991, 1996, 2001, 2006, 20011 and 2016
when the \textit{Sociolinguistic Survey} was made), Municipality, Bilinguals
(Euskaldunak), Passive Bilinguals (Ia Euskaldunak), Monolinguals
(Erdaldunak); then come the percentages of Bilinguals, Passive Bilinguals,
and Monolinguals. Year (1993, 1997 2001, 2006, 2011 when the \textit{Street
Surveys} were made), Street Use of Basque (Euskera Erabilera). In the 2016
Excel file the headings are: Municipality (Udalerria), Year\ of the Survey
(Neurketa), Bilingual (Gaitasuna), Sample Size (Lagina), Street Use of
basque (Euskera Erabilera).

\subsection{Matrix of Expected Pay-offs and Replicator Equation}

The matrix of expected pay-offs attached to the LUG is the following:

\begin{center}
\vspace{-1.0cm} 
\begin{tabular}{ccc}
& $\mathbf{R}$ & $\mathbf{H}$ \\ \cline{2-3}
$\mathbf{R}$ & \multicolumn{1}{|c}{${\alpha (m-n)-c(1-\alpha )},$ ${\alpha
(m-n)-c(1-\alpha )}$} & \multicolumn{1}{|c|}{${\alpha (m-n)-c(1-\alpha )},$ $%
{\ \alpha (m-n)}$} \\ \cline{2-3}
$\mathbf{H}$ & \multicolumn{1}{|c}{${\alpha (m-n)}${\ , }$\ \ {\alpha
(m-n)-c(1-\alpha )}$} & \multicolumn{1}{|c|}{$0$, $\ 0$} \\ \cline{2-3}
\end{tabular}
\end{center}

This is a symmetric matrix. Let $x$ be the proportion of bilinguals playing
strategy $R$. To simplify notation, let $\mathbf{h}=$ ${\alpha
(m-n)-c(1-\alpha )}$, and $\mathbf{j}={\alpha (m-n)}$. Then \textbf{A}
denotes the matrix of pay-offs to the bilingual player in the row position: 
\begin{equation*}
\mathbf{A}=\left[ 
\begin{array}{cc}
\mathbf{h} & \mathbf{h} \\ 
\mathbf{j} & 0%
\end{array}%
\right]
\end{equation*}%
Since we are dealing with a symmetric two player game, the pay-off matrix to
the bilingual player in the column position, denoted \textbf{B}, will be the
transpose of matrix \textbf{A}: \textbf{B = A}$^{\mathbf{T}}$. Members of
the bilingual population are divided in two types, those who choose pure
strategy $R,$ and those who choose $H$\textbf{, }with proportions $x$ and $%
1-x$, respectively. The success of a bilingual type is a function of the
population state $\mathbf{x}=(x,1-x)$. Since the LUG is played continuously
by pairwise random matches, we may assume that the state $\mathbf{x}(t)$
evolves as a differentiable function of time $t$. The rate of increase $%
\overset{\bullet }{x}/x$ of the type \textit{R} depends on the expected
pay-offs to strategy $R$ ( that is, $\mathbf{Ax}$) relative to the average
expected pay-offs (that is, $\mathbf{x}^{\mathbf{T}}\mathbf{Ax}$). Thus, we
obtain the one-population replicator dynamics equation associated to the
LUG, as follows, 
\begin{equation*}
\frac{\overset{\bullet }{x}}{x}=\mathbf{Ax}-\mathbf{x}^{\mathbf{T}}\mathbf{Ax%
}
\end{equation*}%
Clearly $\frac{\overset{\bullet }{x}}{x}>0$ if $\mathbf{Ax}>\mathbf{x}^{%
\mathbf{T}}\mathbf{Ax}$, $\frac{\overset{\bullet }{x}}{x}<0$ if $\mathbf{Ax}%
<x^{\mathbf{T}}\mathbf{Ax}$ , and $\frac{\overset{\bullet }{x}}{x}=0$,
otherwise. The bilinguals who choose $H$ will evolve in the opposite
direction to that of $R$, so there is no need to indicate the differential
equation of $1-$ $\overset{\bullet }{x}$. Doing the operations shown in the
replicator equation, we obtain the differential equation for the evolution
of the share of bilinguals playing pure strategy $R$: 
\begin{equation*}
\overset{\bullet }{x}=x(1-x)[\alpha (m-n)(1-x)-c(1-\alpha )]
\end{equation*}%
The rest points for the equation are $x=1$, $x=0$ and the mixed strategy
equilibrium $x^{\ast }=1-\frac{c(1-\alpha )}{\alpha (m-n)}$. When $%
0<x<x^{\ast }$, $x$ increases toward $x^{\ast }$, and when $1>x>x^{\ast }$, $%
x$ decreases toward $x^{\ast }$. Thus, the asymmetric equilibria ($R,H$) and
($H,R$) are unstable, and $x^{\ast }$ is evolutionary stable.

\textbf{Remarks:} The equilibria of the LUG, in the framework of pairwise
strategic interactions represented as a symmetric two player game, need the
following assumptions:

1. The player role or position in the game does not condition the choice of
action. That is, both the row player (or player I) and the column player (or
player II) will choose freely either $R$ or $H$. On this issue, see Weibull
(1995) and the quoted references therein.

2. Who is the speaker (or who initiates the conversation), and who is the
listener (or hearer) is not conditioned by the player position nor by the
choice of strategy. In particular, the row (column) player is not
necessarily the speaker (listener). An $R$ chooser (either row or column
player) may lead the language $B$ coordination process either as speaker or
as listener. An $H$ chooser is passive in that coordination.

\subsection{The Parameter Estimates}

\begin{table}[tbh]
\caption{Parameter estimates (with standard deviations in parenthesis) and $%
95\%$ bootstrap confidence intervals (C.I.) for the Basque municipals; $obs.$
indicates number of observations.}
\label{tab1}
\begin{center}
\begin{tabular}{l|llllll}
{year} & 1993 & 1997 & 2001 & 2006 & 2011 & 2016 \\ \hline\hline
$\beta_1$ & .862 (.045) & .890 (.027) & .946 (.024) & 1.07 (.026) & .849
(.036) & 1.10 (.024) \\ 
C.I. & [.778;.905] & [.857;.929] & [.918;1.02] & [1.03;1.13] & [.776;.934] & 
[1.09;1.17] \\[1mm] 
$\beta_2$ & .034 (.018) & .030 (.015) & .003 (.014) & .039 (.036) & -.191
(.031) & .008 (.009) \\ 
C.I. & [-.002;.070] & [.000;.060] & [-.019;.035] & [-.021;.116] & 
[-.233;-.116] & [.000;.033] \\[1mm] 
$\beta_3$ & .082 (.039) & .089 (.024) & .011 (.011) & .032 (.023) & .013
(.023) & .001 (.005) \\ 
C.I. & [.043;.155] & [.055;.116] & [.000;.036] & [.001;.096] & [.001;.088] & 
[.000;.015] \\ \hline\hline
obs. & 95 & 116 & 127 & 53 & 74 & 175%
\end{tabular}%
\end{center}
\end{table}
\begin{table}[tbh]
\caption{Parameter estimates (with standard deviations in parenthesis) and $%
95\%$ bootstrap confidence intervals (C.I.) for the Irish local election
areas; $obs.$ indicates number of observations.}
\label{tab2}
\begin{center}
\begin{tabular}{l|lll}
year & 2002 & 2006 & 2011 \\ \hline\hline
$\beta_1$ & .043 (.012) & .007 (.375) & .020 (.144) \\ 
C.I. & [.042;.082] & [.007;1.41] & [.016;.513] \\[1mm] 
$\beta_2$ & -1.31 (.080) & -1.42 (1.28) & -1.02 (.802) \\ 
C.I. & [-1.32;-1.05] & [-1.509;3.04] & [-1.15;2.27] \\[1mm] 
$\beta_3$ & .237 (.043) & .317 (.103) & .100 (.093) \\ 
C.I. & [.101;.245] & [.035;.514] & [.009;.472] \\ \hline\hline
obs. & 180 & 180 & 200%
\end{tabular}%
\end{center}
\end{table}
\begin{table}[tbh]
\caption{Parameter estimates (with standard deviations in parenthesis) and $%
95\%$ bootstrap confidence intervals (C.I.) for the Welsh local authorities.
Estimates are based on $22$ observations in 2004-06.}
\label{tab3}
\begin{center}
\begin{tabular}{l|ll|ll}
& \multicolumn{2}{c}{2004-2006} & \multicolumn{2}{c}{2013-2015} \\ 
& estimate (s.e.) & 95\% C.I. & estimate (s.e.) & 95\% C.I. \\ \hline\hline
$\beta_1$ & .461 (.046) & [.335;.527] & .429 (.019) & [.391;.475] \\ 
$\beta_2$ & -.538 (.048) & [-.629;-.425] & -.667 (.025) & [-.691;-.579] \\ 
$\beta_3$ & .004 (.034) & [.000;.100] & .004 (.007) & [.0003;.027] \\ 
\hline\hline
\end{tabular}%
\end{center}
\end{table}

\subsection{Nonparametric Estimation procedure}

For the ease of notation we always use $KE$ as observed response variable;
for $DU$, the methodology works exactly the same way. Given a sample $%
\{\alpha _{i},KE_{i}\}_{i=1}^{n}$ one wants to estimate the conditional
expectation $E[KE|\alpha ]=g(\alpha )$ under the assumption that $g(\cdot )$
is a smooth function having second order Lipschitz continuous derivatives.
The errors $v=KE-g(\alpha )$ have finite variance. One may add some
conditions on the distribution of $\alpha $ if one wants to calculate the
statistical properties of the now described estimator: For a weight or
kernel function $K(\cdot )$ for which we chose the Epanechnikov kernel $%
K(u)=0.75\cdot (1-u^{2})_{+}$ (the subindex $+$ indicates that the function
is set to zero if $1-u^{2}$ is negative) and bandwidth $h_{x}$ we take 
\begin{equation}
\widehat{g(x)}=\ \mathrel{\mathop{argmin}\limits_{g,g_1}}\sum_{j=1}^{n}%
\left( KE_{j}-g-g_{1}\cdot (\alpha _{j}-x)\right) ^{2}K(\frac{\alpha _{j}-x}{%
h_{x}})  \label{nonparest}
\end{equation}%
as an estimate for $g(x)$. This is the local linear kernel estimator.
Letting $x$ run over the range of $\alpha $ (over all sample observations $%
\alpha _{i}$) we can draw than the function estimate of $g(\cdot )$ which is
compared with our model for $PKE$.

\end{document}